\documentclass[aps,prd,reprint,nofootinbib,showkeys,showpacs,longbibliography,floatfix,preprintnumbers]{revtex4-1}

\pdfoutput=1
\usepackage{amsmath}        % For math environments like align and math symbols
\usepackage{booktabs}       % For professional-looking tables (optional but recommended)
\usepackage{array}          % For column alignment customization
\usepackage{multirow}       % For multi-row cells in tables
\usepackage{graphicx}       % For adjusting table dimensions if needed
\usepackage{dcolumn}        % For decimal-aligned columns
\usepackage{xspace}         % For proper spacing after macros
\usepackage{siunitx}        % For numerical formatting (optional, use if needed)
\usepackage{amssymb}
\usepackage[american]{babel}
\usepackage{booktabs}
\usepackage[T1]{fontenc}  % Font encoding that enables acents in the PDF
\usepackage[]{hyperref}  % Add hypertext capabilities
\usepackage[utf8]{inputenc}  % Accept the Western European characters set
\usepackage{url}
\usepackage[dvipsnames]{xcolor}
\usepackage{float}
\usepackage{tabularx}
\usepackage{booktabs} % For professional-looking table lines
\renewcommand{\arraystretch}{1.5} % Adjust row height for better readability

\DeclareUnicodeCharacter{2212}{-}

\makeatletter

\newcommand{\Rmnum}[1]{\expandafter\@slowromancap\romannumeral #1@}
\definecolor{greenW}{rgb}{0.0, 0.55, 0.1}

\newcommand{\Neff}{N_{\mathrm{eff}} }

\makeatother

\renewcommand {\arraystretch}{1.3}

\hypersetup{
	citecolor = red,
	linkcolor = blue,
	urlcolor = magenta
}

\begin{document}

%\title{Neutrino Cosmology in 2024}
%\title{Foundational Insights and Cutting-Edge Theories in Neutrino Cosmology \\ \ \\ Roots and Frontiers: Neutrino Cosmology from Fundamentals to Future Perspectives \\ \ \\ From Foundations to Frontiers: Insights and Innovations in Neutrino Cosmology \\ \ \\ 
\title{Status of neutrino cosmology: Standard $\Lambda$CDM, extensions, and tensions}

\author{Helena Garc\'ia Escudero}
\email{garciaeh@uci.edu}

\author{Kevork N.\ Abazajian}
\email{kevork@uci.edu}

\affiliation{Center for Cosmology, Department of Physics and Astronomy,
University of California, Irvine, California 92697-4575, USA}

\preprint{UCI-HEP-TR-2024-20}

\begin{abstract}

We examine the performance of the six-parameter $\Lambda$CDM model and its extensions in light of recent cosmological observations, with particular focus on neutrino properties inferred from cosmology. Using a broad suite of nine combinations of datasets, with three separate analyses of the Planck Cosmic Microwave Background (CMB) data, and three separate supernovae (SNe) survey data, plus the recent DESI baryon acoustic oscillation (BAO) scale results, we derive constraints on the sum of neutrino masses ($\Sigma m_\nu$). Our results show upper limits in the range of $\Sigma m_\nu < 76.9\,\mathrm{meV}$ to $\Sigma m_\nu < 108\,\mathrm{meV}$ (95\% CL). The variation in the limits on $\Sigma m_\nu$ arises from the separate analyses of the Planck CMB data and the separate supernovae datasets, as they relate to the inferred matter density and its relation to the sensitivity of the BAO scale and CMB lensing to $\Sigma m_\nu$. In the context of hierarchical mass models in $\Lambda$CDM, we find a $1.47\sigma$ preference for normal ordering (NO) over inverted ordering (IO), with similar values of preference across all datasets. Despite the strong constraints, an inclination towards the non-standard massless neutrinos over NO remains weak at $1.36\sigma$. We find that a ``negative'' neutrino mass, inferred from the shape of the likelihood in the physical regime, $\Sigma m_\nu > 0$, is only present at less than $2\sigma$. The origin of the strong $\Sigma m_\nu$ constraints arise primarily from the high CMB lensing signal, which disfavors the suppression of power from $\Sigma m_\nu$, and the DESI BAO scale, which are complementary to that from the suppression of power.  We confirm that models allowing extra relativistic degrees of freedom, with $N_{\rm eff} \approx 3.5$, alleviate the Hubble tension. Significantly, we find a $3.3\sigma$ preference for a 0.1 eV partially thermalized sterile neutrino when the SH0ES $H_0$ measurement is included, a scale of interest in short-baseline oscillation experiment results. When $H_0$ is included, fully thermalized sterile neutrino models are as consistent as $\Lambda$CDM in fitting all datasets, but are disfavored otherwise. We also explore an 11-parameter model relaxing the dark energy equation of state and curvature, together with  $\Sigma m_\nu$ and $N_{\rm eff}$,  finding consistency with $\Lambda$CDM parameters except for the dark energy equation of state $w_0=-0.961^{+0.012}_{-0.037}$. Neutrino mass constraints in this extended model remain stringent, with $\Sigma m_\nu < 97.0\,\mathrm{meV}$ (95\% CL). 

\end{abstract}
%----------------------------------------------
\maketitle
%------------------------------------------------

\section{Introduction}\label{sec:intro}
Neutrinos are among the most enigmatic particles in the Standard Model (SM) of elementary particle physics. However, they are vital in cosmology, as they impact the Universe's evolution and contribute to the formation of the large-scale structures we see today. Their properties and nature can be studied by the detectable imprints they leave on cosmological data. When neutrinos were definitively measured to oscillate and have mass over 25 years ago, their impact on cosmology was understood to be significant (for reviews, see Refs.~\cite{Lesgourgues:2006nd,Hannestad:2010kz,Abazajian:2016hbv}). 
The 6-parameter cosmological standard model has been exceedingly successful in its consistency with observational data from the last-scattering surface of the cosmic microwave background (CMB) to galactic scales today \cite{Planck:2018vyg}. The principle constituents of this model are cold dark matter and dark energy in the form of a cosmological constant, so it is referred to as $\Lambda$CDM. Neutrinos with arbitrary mass are an extension of $\Lambda$CDM, since their oscillations measure the difference of the squares of the neutrinos' mass, and not their absolute mass scale. In addition, if neutrinos are non-thermal or have extra states, this adds another measurable unknown to minimal $\Lambda$CDM. 

Currently, cosmological data are already offering constraints on neutrino properties that are not only complementary to terrestrial experiments but also competitive with them. For example, leading cosmological results on neutrino mass are approximately a factor of 20 more stringent \cite{DESI:2024mwx}, as cosmology measures the sum of neutrino mass eigenstates, $\Sigma m_\nu$, while direct kinematic laboratory measurements via nuclear $\beta$-decay are sensitive to the effective electron antineutrino mass \cite{Katrin:2024tvg}. Cosmological constraints on neutrino properties are model-dependent in that they require the minimalist assumption of a single-parameter extension to $\Lambda$CDM, $\Sigma m_\nu$.
The existence of a thermal relic cosmic neutrino background is a fundamental prediction of the standard hot Big Bang model and $\Lambda$CDM. Though it has not been directly detected, indirect evidence supports its presence through the primordial abundance of light elements produced during Big Bang Nucleosynthesis (BBN), the power spectrum of CMB anisotropies, and cosmological large-scale structure \cite{Cyburt:2015mya}. These observations would be significantly inconsistent without a relativistic cosmological component that closely matches the properties predicted by the standard neutrino decoupling process, involving only weak interactions. 
As a result, cosmology is attuned to specific neutrino properties: their density, linked to the number of active neutrino species, and their masses. 
$\Lambda$CDM assumes that the only light, relativistic relic particles since BBN are photons and active neutrinos. Extended models propose additional light particles, such as sterile neutrinos or thermal axions, which can mimic the effects of active neutrinos and complicate the interpretation of the inferred relic neutrino density, since new relativistic components can both add to the relativistic energy density, or replace that of the active neutrinos. These models have been explored to address anomalies like the Hubble tension, the discrepancy between inferred and direct measurements of the Universe's expansion rate. 

In the $\Lambda$CDM model, three active neutrinos thermalize in the early Universe, decoupling gradually around temperatures of 2 MeV. The effective number of neutrino species, $N_{\textrm{eff}}$, is defined via
\begin{equation}
    \frac{\rho_\mathrm{rad}}{\rho_{\gamma}} = \frac{7}{8} N_{\textrm{eff}}\left(\frac{4}{11}\right)^{4/3}\, .
\end{equation}
Here, $\rho_\mathrm{rad}$ and $\rho_{\gamma}$ represent the energy density of non-photon radiation and photons respectively. $N_{\textrm{eff}}$ quantifies the relativistic energy density beyond photons, which could be any relativistic relic component. In the standard cosmology, $N_{\textrm{eff}}$ is composed solely by the active neutrino density contribution, i.e., $N_{\textrm{eff}} = N_\nu^\mathrm{SM}$, with precise calculations predicting $N_\nu^\mathrm{SM} = 3.044$ when accounting for neutrino oscillations, incomplete neutrino decoupling through $e^\pm$ annihilation, and QED corrections \cite{Akita:2020szl,Bennett:2020zkv,Froustey:2020mcq}. Cosmological observations' confirmation of this value is a crucial test of the standard $\Lambda$CDM model, and observations that definitively infer a value deviating from $N_\nu$ would be a major breakthrough. In concert, the adoption of $N_{\textrm{eff}} = 3.044$ for interpreting the effects of the standard active neutrinos is an essential standard model assumption to place neutrino mass limits with cosmology.

An increased $N_{\textrm{eff}}$ influences the observable power spectrum of CMB anisotropies and matter clustering through changes in the background cosmological expansion through matter-radiation equality, and changes in the growth of perturbations. If the densities of other components are held constant, a higher $N_{\textrm{eff}} $ delays radiation-to-matter equality, leading to significant alterations in the CMB anisotropies, specifically suppressing the high-$\ell$ multipole peaks relative to their position in $\ell$. To better isolate the effect of  $N_{\textrm{eff}} $, it is useful to increase the densities of total radiation and matter uniformly, keeping the redshift of matter-radiation equality fixed. This adjustment reveals that a higher  $N_{\textrm{eff}} $ increases the diffusion (Silk damping) scale at decoupling, resulting in decreased CMB temperature power at high multipoles relative to the sound horizon, which determines the acoustic peak positions \cite{Abazajian:2016hbv}. The free-streaming radiation parameterized by $N_{\textrm{eff}}$ also produces a phase-shift in the acoustic peaks that is unique \cite{Bashinsky:2003tk}, and has been detected \cite{Baumann:2015rya}. 

Additionally, neutrino's dual behavior, acting as radiation at early times and transitioning to dark matter-like evolution during later structure formation, places them in an unique position in cosmology. As neutrinos transition from relativistic to non-relativistic states, they initially travel at nearly the speed of light, which affects their diffusion scale and has important consequences for gravitational clustering and the growth of large-scale structures. Once they become non-relativistic, the energy density of neutrinos is primarily determined by their total mass:
\begin{equation}
    \Omega_{\nu} = \frac{\rho_{\nu}^0}{\rho_{\rm crit}^0} = \frac{\Sigma m_{\nu}}{93.13 ~ h^2 ~ \rm eV}\, ,
    \label{eq:omnu}
\end{equation}
where $\rho_{\nu}^0$ and $\rho_{\rm crit}^0$ are respectively the total neutrino and critical density of the Universe today and $h$ is the Hubble constant in units of $100\,\mathrm{km \,s^{−1}\,Mpc^{−1}}$. In this expression, we use the temperature of neutrinos that provides $N_\nu = 3.044$. The absolute neutrino mass scale has been probed using beta decay, with the KATRIN experiment currently providing the best bound on the beta-decay effective neutrino mass of $m_{\nu,\beta} < 0.45\,\text{eV}$ at 90\% confidence \cite{Katrin:2024tvg}. %Despite these advancements, this bound is still about 20 times less stringent than state-of-the-art cosmology constraints \cite{DESI:2024mwx}.

The presence of massive neutrinos has varied effects on CMB observables. An increase in neutrino mass impacts the late-time non-relativistic matter density, altering the angular diameter distance to recombination and the late integrated Sachs Wolfe (ISW) effect. The transition of neutrinos to a non-relativistic state also induces variations in the Universe's pressure-to-density ratio, observable through the early ISW effect, and reduces the overall signal of weak lensing of CMB anisotropies because of free-streaming suppression of the large scale structure that produces the lensing of the CMB. It is this last observable, lensing of the CMB, that makes high angular resolution CMB observations highly sensitive to neutrino mass \cite{Kaplinghat:2003bh}. Because the suppression of the matter power spectrum of large-scale structure (LSS) by neutrinos occurs below the horizon scale of matter-radiation equality, an increase in the dark matter density, $\Omega_\textrm{cdm}$, can compensate the presence of matter neutrinos in LSS clustering observations, making neutrino mass positively correlated with increases in $\Omega_\textrm{cdm}$.

The baryon acoustic oscillation (BAO) signature has become a standard cosmological signal measurable by several cosmological surveys \cite{Weinberg:2013agg}. The BAO peak in real space, measured through the correlation function of matter tracers such as galaxies and gas, is imprinted by the expansion of overdensities moving at the sound speed of the plasma at the time, until the moment of photon decoupling. The length of this motion is the sound horizon. The sound horizon, $r_s$ at the decoupling redshift, $z_d$, is approximately given by
\begin{equation}
    r_\mathrm{s}(z_\mathrm{d}) = \int^{z_\mathrm{d}}_\infty {\frac{c_s\, dz}{H(z)}}
\end{equation}
where $c_s$ is the sound speed, and $H(z)= H_0\sqrt{\Omega_m(1+z)^3+\Omega_r(1+z)^4}$ is the Hubble rate through the early to decoupling epoch. Here, $\Omega_m$ and $\Omega_r$ are the critical densities in matter and radiation, respectively. Increases in both $\Omega_m$ and $N_{\textrm{eff}}$ (via $\Omega_r$) both increase their contribution to $H(z)$, and decrease the sound horizon and the BAO scale, which makes BAO observations very sensitive to the matter and radiation content of the early Universe. The cosmological evolution of the BAO scale provides it as a powerful standard ruler for measurements of dark energy and its evolution \cite{Seo:2003pu}. Neutrino mass affects the BAO length-scale when neutrino mass is sufficiently large as to become non-relativistic during the drag epoch, before photon decoupling. In linear theory, the BAO scale diminishes with increasing neutrino mass because, though the neutrinos free stream and redshift as radiation in early times, they augment $\Omega_m$ during the BAO drag epoch, reducing $r_s$ as an increase in $\Omega_m$ would. Therefore, the BAO scale generally makes $\Sigma m_\nu$ anti-correlated with $\Omega_\textrm{cdm}$, complementary to their degeneracy with LSS clustering  \cite{Thepsuriya:2014zda}. Because the geometrical information in the BAO scale comes in the form of a ratio of $r_s$ to the angular diameter and line-of-sight distances, the geometry measured by the BAO scale is both sensitive to the neutrino mass, and can also change the relation of $\Sigma m_\nu$ and $\Omega_\textrm{cdm}$  from anti-correlation to correlation as distances' dependencies dominate over $r_s$ dependencies \cite{BOSS:2014hhw}. As detailed in Ref.~\cite{Loverde:2024nfi}, geometric information from the BAO scale and its ratio to distances can be as sensitive---or even more-so---than growth of structure probes of neutrino mass.

Supernovae (SNe) observations measure the relationship between the luminosity distance and redshift in the local Universe. This is dependent on the late-time cosmological energy density, and therefore, SNe are not directly sensitive to neutrino mass or radiation energy density. However, SNe do constrain parameters degenerate with neutrino mass in CMB and BAO observables, primarily $\Omega_m$ and the dark energy equation of state (EoS), $w$. We include SNe because of their complementary sensitivity to CMB and BAO determinations of cosmological growth and expansion history. In Ref.~\cite{Loverde:2024nfi}, it was shown that middle to low-redshift geometrical measures from BAO and SNe are highly sensitive to the total cosmological matter content as a \textit{fraction} of the critical density, as opposed to the \textit{physical} density, and therefore are highly sensitive to the neutrino mass contribution to this fractional density. The sensitivity is such that fixing the angular sound horizon from the CMB gives low-redshift measures a scaling of $\Omega_m \propto (1+f_\nu)^5$, where $f_\nu \equiv \Omega_\nu/\Omega_m$.

Much of our work studies the constraints on neutrinos as minimal extensions beyond the 6-parameter framework of $\Lambda$CDM, which successfully describes nearly all cosmological observables with the amplitude of scalar perturbations, $A_s$, the spectral index of these perturbations, $n_s$, the optical depth to the last-scattering surface of the CMB, $\tau$, an angular scale corresponding to the sound horizon of the last scattering surface, $\theta_\mathrm{MC}$, as well as the density of cold dark matter and baryons relative to the critical density, $\Omega_\textrm{cdm}$ and $\Omega_\textrm{b}$. We also study an 11-parameter extension of $\Lambda$CDM which allows variation of the evolution of the dark energy density through $w_0$ \& $w_a$, the curvature of the Universe, $\Omega_k$, as well as the neutrino-related parameters, $\Sigma m_\nu$ \& $\Neff$.

The structure of this paper is as follows: Section \ref{sec:lik} details the datasets and likelihoods adopted in our analysis. Section \ref{sec:meth} explains the statistical methods we employed. Section \ref{sec:LCDM} presents our findings within $\Lambda$CDM, including constraints on $\Sigma m_\nu$ and neutrino mass ordering, as well as $N_{\textrm{eff}}$. In Section \ref{sec:model}, we discuss tension datasets, including $H_0$ and $S_8$, and the neutrino-related extensions of $\Lambda$CDM that the tensions motivate, including massive neutrinos and extra (sterile) neutrinos. Finally, we conclude and discuss the future outlook in Section \ref{sec:concl}.

\section{Cosmological Datasets}
\label{sec:lik}

We present here the cosmological datasets and methodology adopted in our work. We selected those datasets that are among the most robust measures of cosmological growth and expansion history for their epochs, and which span a wide range of cosmological history. 
%due to their robustness and comprehensiveness. 
That is, they are those observations that have the minimum statistical errors for their measurements. In addition, these datasets have undergone thorough testing for systematic effects, as we discuss below. Lastly, these observations provide strong leverage for tests of cosmology with measurements of expansion and growth history over the widest possible observational range, from the surface of last scattering to low redshifts.

\begin{table}[t]
\centering
\begin{ruledtabular}
\begin{tabular}{lcll}
\multicolumn{4}{c}{Datasets} \\ \hline
CMB Planck          & CMB Lensing                                                         & BAO              & SNe               \\ \hline
P18                 & \multirow{3}{*}{\begin{tabular}[c]{@{}c@{}}Lensing PR4 + \\ ACT DR6\end{tabular}} & \multirow{3}{*}{DESI DR1} & PP                \\ 
CamSpec             &                                                                     &                   & U3                \\ 
P20                 &                                                                     &                   & DES Y5            \\ 
\end{tabular}
\end{ruledtabular}
\caption{Summary of the cosmological datasets we considered in this work, as introduced in Section~\ref{sec:lik}.}
\label{table:Datasets}
\end{table}

\subsection{Planck CMB}
The Planck survey of CMB  has precisely measured the CMB anisotropies via their temperature and polarization angular power spectra, with five out of six $\Lambda$CDM parameters determined to better than 1\% with Planck data alone, with the most accurately known parameter, the angular sound horizon, now measured to 0.03\%. Over the years, Planck data and its analysis has advanced considerably, starting with the initial data release in 2013, followed by the incorporation of polarization data in 2015, and culminating in significant enhancements to systematic corrections in the ultimate 2018 data release. Since that final data release, there have been two independent reanalyses of the Planck data products. We consider all three:\\

\noindent \textbf{Planck 2018 (P18)}:  we use CMB temperature and polarization anisotropy power spectra (and their cross-spectra) from the Planck 2018 legacy data release (PR3)~\cite{Planck:2018vyg,Planck:2019nip}. Specifically, we use: the high-$\ell$ \texttt{Plik}; the low-$\ell$ \texttt{commander} likelihood for the TT spectrum in the multipole range $2 \leq \ell \leq 29$; likelihood or the TT spectrum in the multipole range $30 \leq \ell \leq 2508$ and for the TE and EE spectra in the range $30 \leq \ell \leq 1996$, and the low-$\ell$ \texttt{SimAll} likelihood for the EE spectrum in the range $2 \leq \ell \leq 29$. It is worth mentioning that the lensing anomaly for this analysis is at a $2.8\sigma$ level and significantly reduced in the two reanalysis we introduce next \cite{Naredo-Tuero:2024sgf}.\\
%LE

\noindent\textbf{CamSpec:} CamSpec project reanalyzed the Planck temperature and polarization maps of the CMB of each Planck data release. The motivation for creating this new revised likelihood was to provide a single-source detailed description of the construction of likelihoods for the Planck collaboration as well as testing the consistency and fidelity of the Planck data. CamSpec included additional methods for improved subtraction of Galactic dust emission, extending sky coverage and improving accuracy in temperature and polarization power spectra, compared to previous Planck analyses \cite{Efstathiou:2019mdh}.  The lensing anomaly in this analysis is reduced when compared to P18 and it is at a  $1.7\sigma$ level \cite{Naredo-Tuero:2024sgf}.\\

\noindent \textbf{Planck NPIPE (P20)}:  the Planck Collaboration released a new processing pipeline in 2020 for generating calibrated frequency maps from the final Planck data. This latest reprocessing of data from both the low-frequency (LFI) and high-frequency (HFI) instruments, using the integrated NPIPE pipeline, resulted in more data, reduced noise, and improved consistency across frequency channels \cite{Planck:2020olo}. HiLLiPop and LoLLiPoP are the new likelihoods of the CMB temperature and E-mode polarization power spectra from this pipeline, covering both large scales (low-$\ell$) and small scales (high-$\ell$). These reanalysis uses a higher sky fraction (75\%), as well as a wider range of multipoles and a negligible modeling dependence on cosmological $\Lambda$CDM parameters. As a result, the constraints derived have 10\% to 20\% smaller uncertainties on $\Lambda$CDM parameters compared to previous pipelines. We employ the HiLLiPop and LoLLiPoP likelihoods obtained from this reanalysis of the CMB temperature and polarization power spectra for the high-$\ell$ ($30<\ell<2500$) TT, TE, and EE data \cite{Tristram:2023haj}. For the low-$\ell$ ($\ell<30$) EE spectra, we utilize the LoLLiPoP likelihoods \cite{Tristram:2023haj}. This analysis has the smallest level of the lensing anomaly of all the CMB analysis we include, it is at a $0.75\sigma$ level \cite{Naredo-Tuero:2024sgf}. We term the combination of these likelihoods as ``\textbf{P20.}''

\subsection{CMB lensing}

CMB surveys also capture the power spectrum of the gravitational lensing potential, $C_l^{\phi \phi}$, via 4-point correlation functions.
The Planck PR4 lensing map utilizes a quadratic estimator pipeline on CMB maps derived from the enhanced NPIPE reprocessing of data from the Planck HFI. This version incorporates approximately 8\% more CMB data compared to Planck PR3, which includes information obtained during satellite re-pointing maneuvers, along with several enhancements in data processing \cite{Planck:2020olo}.
We additionally consider the new measurements of CMB lensing from the Atacama Cosmology Telescope (ACT) Data Release 6 (DR6) covering over 9,400 square degrees of the sky. These lensing measurements, based on five seasons of CMB observations from ACT, achieve a precision of 2.3\% for the CMB lensing power spectrum. This analysis pipeline was developed to reduce sensitivity to foreground contamination and noise characteristics, and included a number of systematic error assessments~\cite{ACT:2023kun}.  
We utilize the latest NPIPE PR4 Planck CMB lensing reconstruction \cite{Carron:2022eyg} alongside DR6 from the ACT \cite{ACT:2023dou}. For brevity, this dataset combination is referred to as ``\textbf{CMB lensing}.''

\subsection{BAO}
%\textbf{DESI Data Release (DR1)}: 
Dark Energy Spectroscopic Instrument's (DESI) BAO measurements provide robust data on the transverse comoving distance and the Hubble rate, or their combination relative to the sound horizon, across seven redshift bins derived from over 6 million extragalactic objects in the redshift range of $ 0.1 < z < 4.2$. We employ the latest BAO scale measurements from the DESI Data Release 1 \cite{DESI:2024mwx}. This dataset includes the Bright Galaxy Sample (BGS: $0.1 < z < 0.4$), Luminous Red Galaxy Sample (LRG: $0.4 < z < 0.6$ and $0.6 < z < 0.8$), Emission Line Galaxy Sample (ELG: $1.1 < z < 1.6$), the combined LRG and ELG Sample (LRG+ELG: $0.8 < z < 1.1$), Quasar Sample (QSO: $0.8 < z < 2.1$), and the Lyman-$\alpha$ Forest Sample (Ly$\alpha$: $1.77 < z < 4.16$). We refer to the full DESI dataset is referred to as ``\textbf{DESI}.''

DESI offers several advancements over previous BAO analyses, including more physically-motivated enhancements to BAO fitting and reconstruction methods.  DESI implements blind analysis techniques to reduce confirmation bias. Furthermore, DESI has a factor of two to three improvement in survey volume and number of tracers relative to SDSS-IV \cite{eBOSS:2020yzd, Smith:2020stf}. It also applies a unified BAO analysis approach across different tracers and allows for a combination of these tracers' measure of the BAO scale \cite{DESI:2024uvr}.   

\subsection{Type Ia SNe}
Type Ia SNe (SNe Ia) are crucial standardizable candles that offer an alternative approach to measuring the Universe's expansion history. While SNe measurements have lower statistical power on the expansion history compared to modern  BAO measurements within the $\Lambda$CDM model, they provide essential information on dark energy, particularly when examining less restricted cosmological models. In our analysis, we include the following three SNe datasets:

\smallskip 
\noindent \textbf{Pantheon+ (PP)}: We utilize the updated SNe Ia luminosity distance measurements from the Pantheon+ (PP) sample \cite{Scolnic:2021amr,Brout:2022vxf}, comprising 1550 spectroscopically-confirmed SNe Ia in the redshift range $0.001 < z < 2.26$. The PP sample consists of light curves from these verified SNe Ia, which are analyzed to derive cosmological parameters. The different datasets, with all relevant data, references, and their respective redshift ranges outlined in Table 1 of \cite{Scolnic:2021amr}. Compared to the previous Pantheon dataset \cite{Pan-STARRS1:2017jku}, the PP catalog includes improved photometric calibration and systematic uncertainties. The PP catalog shows the most significant increase in SNe Ia data at lower redshifts, largely due to the addition of the LOSS1, LOSS2, SOUSA, and CNIa0.2 SNes. The PP likelihood includes statistical covariance and models systematic uncertainties as covariances.  

\smallskip
\noindent \textbf{Union3 (U3)}: The Union3 Supernova compilation is an extensive dataset of 2,087 spectroscopically-confirmed SNe Ia, with 1,363 in common with the PP dataset \cite{Rubin:2023ovl}, standardized on a consistent distance scale through SALT3 light-curve fitting.  It aims to manage systematic uncertainties that arise as the number of usable supernovae increases. The dataset addresses outliers and selection biases. They incorporate a Bayesian framework, called UNITY1.5 (Unified Nonlinear Inference for Type-Ia cosmologY), that improves modeling of selection effects, standardization, and systematic uncertainties compared to traditional simulation-based methods.  

\smallskip
\noindent \textbf{Dark Energy Survey Year 5 (DES Y5)}: The DES Y5 sample is the largest and most comprehensive single-campaign SNe Ia survey to date, totaling 1,829 SNe: 1,635 from DES measurements and 194 from local SNe Ia Ne measurements from the CfA/CSP foundation sample \cite{DES:2024tys}. This dataset represents observations gathered during the first five years of the DES Supernova Program. The SNe cover a redshift range of  $0.01 < z < 1.3$ , yielding a substantial dataset for cosmological studies. The DES Y5 sample offers improved high-redshift statistics compared to the PP or U3 datasets.  

It is important to note that the PP, U3 and DES Y5 datasets contain supernovae that overlap between the samples. To avoid any double counting, these datasets are never utilized together. Table \ref{table:Datasets} summarizes all the datasets that we have just defined.
For conciseness, we use the term ``\textbf{benchmark}'' to refer to the combination of the likelihoods from P18, CMB lensing, DESI DR1, and PP described above.

\subsection{Tension Datasets}
We also consider two measurements of the late Universe that are internally robust but are in strong tension with the cosmology inferred from our datasets above. Specifically, these are: the local Hubble expansion rate, $H_0$ and the amplitude of matter fluctuations, $\sigma_8$, at the scale of $8,\mathrm{Mpc}/h$, sometimes parameterized as $S_8$ ($S_8 = \sigma_8 \sqrt{\Omega_{m}/0.3}$). For the latter, $S_8$, we adopt a recent well-tested measurement. These data are:

\smallskip
\noindent \textbf{SH0ES ($H_0$):} the Gaussian likelihood of the Hubble constant calculated by the measurements of the SH0ES collaboration in~\cite{Riess:2021jrx}, $H_0 = 73.04 \pm 1.04\,{\rm km/s/Mpc}$. The selection of this result is motivated by the reduced error bars obtained by the collaboration to calculate this value compared to the one obtained following other analysis. We employ the explicit value of $H_0$, and its error, in our likelihood analysis, and not the absolute magnitude, $M_b$, as the latter is only important for models that differ from $\Lambda$CDM in their expansion history near $z=0$, e.g., $w$CDM. Since these models do not appreciably change our conclusions with respect to the neutrino physics of interest to this work, we use $H_0$ itself.
    
\smallskip
\noindent\textbf{Dark Energy Survey Year 3 ($S_8$):} The Dark Energy Survey (DES) uses a novel photometric redshift calibration technique for its survey of galaxies. The DES photometric calibration method successfully recovers the input cosmological parameters in simulated surveys by using several statistical quantities derived from  the survey data. The cosmic shear measurements are based on an analysis of over 100 million source galaxies, and they constraint the clustering amplitude to be $S_8 = 0.759^{+0.025}_{-0.023}$ \cite{DES:2021bvc}. Our choice of this value of the $S_8$ parameter is motivated by the several verification tests for potential systematics performed by the DES collaboration.

\begin{table}[t!]
\centering
\renewcommand{\arraystretch}{1.5}
\begin{ruledtabular}
\begin{tabular}{l @{\hspace{2cm}} c}
\toprule
\textbf{Parameter} & \textbf{Prior} \\
\hline \hline
$\Omega_\mathrm{b} h^2$ & $[0.005, 0.1]$ \\
$\Omega_\mathrm{c} h^2$ & $[0.01, 0.99]$ \\
$\log(10^{10} A_\mathrm{s})$ & $[1.61, 3.91]$ \\
$n_\mathrm{s}$ & $[0.8, 1.2]$ \\
$\tau_{\rm reio}$ & $[0.01, 0.8]$ \\
$100\theta_\mathrm{MC}$ & $[0.5, 10]$ \\
\hline \hline
$\Sigma m_{\nu}$ [eV] & $[0, 10]$\\
$N_{\rm eff}$ & $[0, 100]$\\
$\Omega_k$ & $[-0.3, 0.3]$\\
$w_0$ & $[-3, 0.1]$ \\
$w_a$ & $[-3, 3]$ \\
%\hline %\hline
\bottomrule
\end{tabular}
\end{ruledtabular}

\caption{We list our adopted ranges for the flat prior distributions imposed on the 6 $\Lambda$CDM, as well as the 5 additional extended-model cosmological parameters.}
\label{tab-priors}
\end{table}

\section{Methodology} 
\label{sec:meth}
In our work, we will be testing neutrino physics scenarios that can be considered to fit within the $\Lambda$CDM framework (such as non-trivial neutrino mass), as well as tests of models beyond $\Lambda$CDM. We fit the cosmological models we study by using the publicly available Bayesian analysis framework \texttt{cobaya}~\cite{Torrado:2020dgo}, with the Markov chain Monte Carlo (MCMC) sampler~\cite{Lewis:2002ah,Lewis:2013hha} and fastdragging~\cite{2005math......2099N}. We employ the \texttt{CAMB} cosmological Boltzmann solver \cite{Lewis:1999bs,Howlett:2012mh}. We run 4 parallel MCMC chains, and use the Gelman-Rubin convergence test, $R-1 < 0.01$, to verify their convergence for all parameters. In all the runs we choose flat priors with cutoffs well outside where likelihoods are significant. The flat prior ranges within which the parameters are varied are given in Tab.~\ref{tab-priors}.
We use GetDist \cite{Lewis:2019xzd} to calculate Bayesian posteriors, means, and confidence intervals.

For finding the best-fit likelihood points and their $\chi^2$ values we use the minimizer sampler available in \texttt{cobaya}~\cite{2018arXiv180400154C,2018arXiv181211343C,Powell_2009}. For all of the scenarios we consider we maximize the likelihood, $\mathcal{L}$, via minimizing its $- 2 \ln{\mathcal{L}}$,  as we are interested in performing a likelihood ratio for our model-preference test statistic. To identify the true minimum best-fit within an acceptable tolerance, we perform multiple minimizations of  $- 2 \ln{\mathcal{L}}$, until at least six results yield a minimum value within 0.2 of the minimum determined by the sampler. The motivation for these specific choices are statistically motivated. We model the evaluation of the minimum value within 0.2 of its true value as a Poisson process. To have a  99.5\%  confidence that the minimal bin includes an evaluation within 0.2 of the true minimum, one needs at least 5 evaluations within the bin (e.g., see Table II of Ref.~\cite{1986ApJ...303..336G}). Following this method, we expect to have found the $- 2 \ln{\mathcal{L}}$ within 0.2 of its minimum at roughly 99.5\% confidence. The likelihood combinations discussed in this section involve over 20 nuisance and free parameters, making the minimization process quite complex. For some datasets and models, the minimum can be found with approximately 200 evaluations. However, the HiLLiPoP likelihood requires significantly more evaluations, $\gtrsim$2000, to achieve our criteria compared to other likelihoods. This is likely because, as noted by others, HiLLiPoP is the only unbinned CMB likelihood, resulting in higher noise in the likelihood function \cite{Naredo-Tuero:2024sgf}. We decided not to include this likelihood for the results that involved minimizing $\chi ^2$ since it did not fulfill our tolerance requirements without the calculation of an excessive number of minimizations.
Our statistical approach follows that of Ref.~\cite{Escudero:2022rbq}. We employ the minimal chi-square $\chi^2_{\rm min}$ to quantify the success of each scenario with respect to a baseline model, \textit{i.e.}, a $\Delta \chi^2$ test.

We compute the change in the effective minimal chi-square $\chi^2_{\rm min} = - 2 \ln{\mathcal{L}} $, where $\mathcal{L}$ represents the maximum likelihood for the considered model $\mathcal{M}$.
The $\Delta \chi^2_{\mathcal{M}}$ relative to the baseline model $\mathcal{M}_0$ is then derived as
\begin{align}
    \Delta \chi^2_{\mathcal{M}} \equiv \chi ^2_{{\rm min},\mathcal{M}} -\chi^2_{{\rm min},\mathcal{M}_0}\,.
    \label{eq:chi2}
\end{align}
The $\chi^2$ value of a dataset can be used to determine if a trend in the data is happening due to chance or due to a new model component, and can also be used to test a model's ``goodness of fit''~\cite{10.2307/1402731}. 

The $\Delta \chi^2_{\mathcal{M}}$ test does not consider each model's complexity, specifically the number of parameters involved.
To account for model complexity, we also use the Akaike information criterion (AIC) within a Bayesian framework, which enables a fair comparison among models with varying parameter counts.
To evaluate the improvement in fit quality, we calculate the AIC for each model, defined as ${\rm AIC} = -2 \ln{\mathcal{L}} + 2k$, where $k$ represents the number of parameters in the model ~\cite{1100705}.
A model is favored over another if it results in a lower AIC.
To make a comparison with $\Lambda$CDM, we compute the AIC of $\mathcal{M}$ relative to that of $\Lambda$CDM, defined as
\begin{align*}
    \Delta {\rm AIC} \equiv -2 (\ln{\mathcal{L_{\mathcal{M}}}}-\ln{\mathcal{L}_{\Lambda{\rm CDM}}}) + 2(N_{\mathcal{M}} - N_{\Lambda{\rm CDM}} )\,.
\end{align*}
Here, $N_{\mathcal{M}}$ and $N_{\Lambda{\rm CDM}}$ denote the number of free parameters in models $\mathcal{M}$ and $\Lambda{\rm CDM}$, respectively. % 
The AIC penalizes models that add extra parameters without sufficiently enhancing the fit; consequently, a model with a lower AIC value is theoretically and statistically more favorable than one with a higher value. 
To evaluate the performance of each model, we interpret AIC values using Jeffreys' scale~\cite{jeffreys1998theory}. This scale, empirically calibrated, offers descriptive terms for varying levels of evidence. Our criterion for favoring model $\mathcal{M}$ over $\Lambda$CDM is when it falls in the ``strong'' category on Jeffreys' scale, meaning $\Delta \text{AIC} <-6.91$.

In addition, we want to quantify the tension when adding $H_0$ or $S_8$ measurements to our datasets. We calculate
\begin{align}
    \sqrt{ \Delta \chi^2_\mathcal{D}} \equiv  \sqrt{\chi^2_{{\rm min},\mathcal{D+T}} -\chi^2_{{\rm min},\mathcal{D}}} \,.
\label{eq:tension}
\end{align} 
In this equation the subscript ${\mathcal{D}}$ denotes the baseline datasets considered, while ${\mathcal{T}}$ indicates the tension constraints incorporated into the minimization calculation. This test does not evaluate a model's effectiveness in describing the data; instead, it quantifies the level of tension within a given model when additional data is included.

\section{$\Lambda$CDM and Neutrino cosmology}
\label{sec:LCDM}
\begin{figure}[]
\centering
\includegraphics[width=\columnwidth, height=0.73\columnwidth]{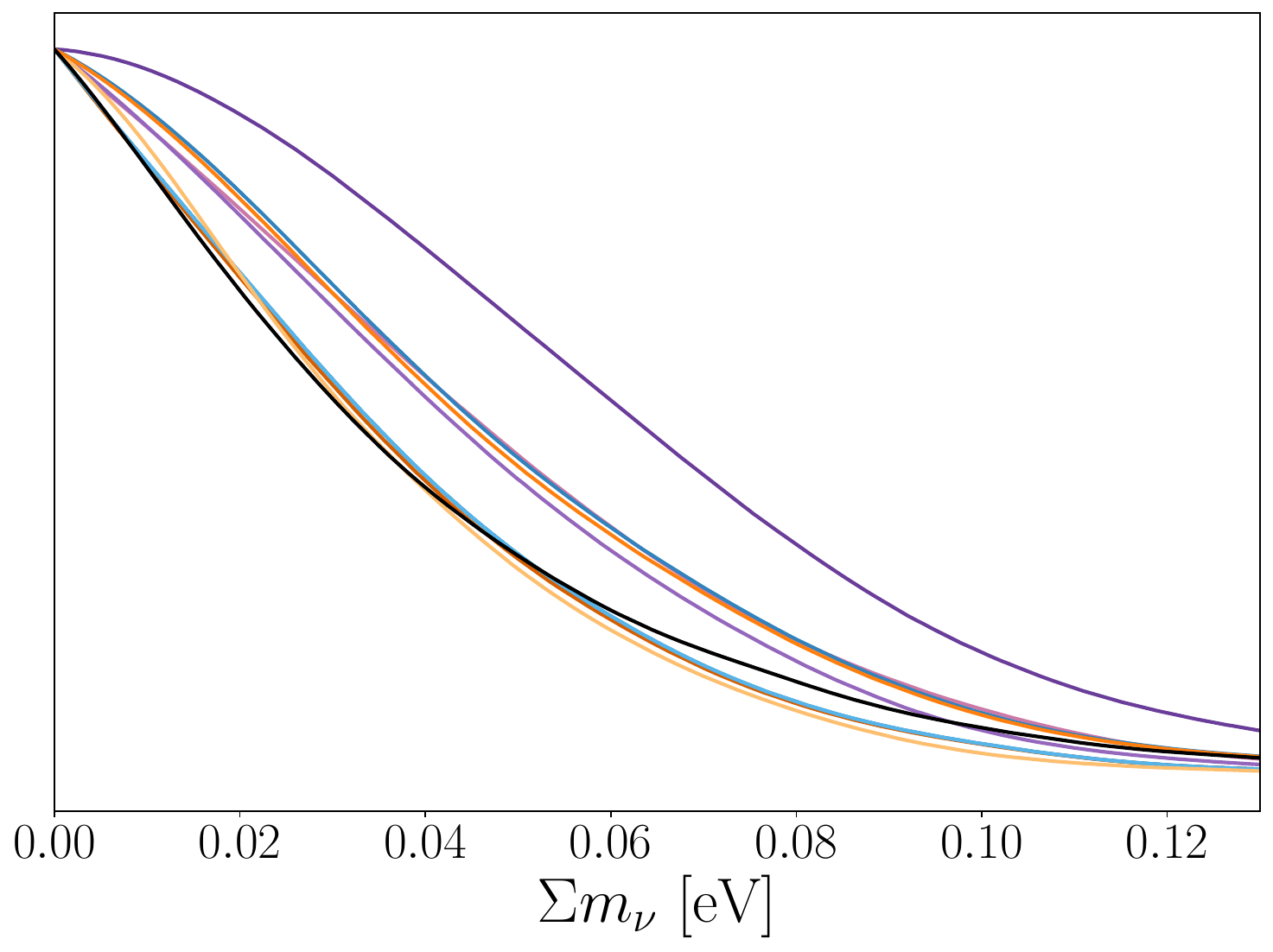}
   \caption{\label{fig:1}Comparison of 1D marginalized posterior distributions for $\Sigma m_{\nu}$ [eV]  for different data combinations. As seen above, there are three clusters of likelihoods that are stringent, moderate, and weakest. The stringent combination includes five likelihoods, with P18, DESI, \& U3 being orange, P18, DESI, \& PP being green, with CamSpec, DESI, \& U3 being red, CamSpec, DESI, \& PP being grey and the 11 parameter chain with our benchmark data in black. The moderate cluster combination includes four datasets' likelihoods, with P18, DESI, \& DES Y5 being red, CamSpec, DESI, \& DES Y5 being light blue, with P20, DESI, \& U3 being dark blue, and P20, DESI, \& PP being green. The weakest likelihood contour, in purple, is for P20, DESI, \& DES Y5. All datasets include CMB lensing. }   
\end{figure}

From an experimental standpoint, cosmology can no longer be sufficiently described by the six-parameter $\Lambda$CDM model assuming massless neutrinos. Solar, atmospheric, accelerator, and neutrino oscillation experiments, such as those conducted by the Super-Kamiokande \cite{Super-Kamiokande:1998kpq,Super-Kamiokande:2001ljr,Super-Kamiokande:2005mbp} , SNO \cite{SNO:2002tuh}, MINOS \cite{SNO:2002tuh}and NOvA \cite{NOvA:2004blv}, and KamLAND \cite{KamLAND:2002uet}, have provided clear evidence for neutrino mass through observed flavor conversion. These laboratory findings confirm that neutrinos have mass and current cosmological data is sensitive to the gravitational effects of their non-relativistic nature. 

However, the observable effects of neutrino oscillations are sensitive only to mass-squared splittings among eigenstates, thus constraining only the neutrino mass-squared differences. Matter effects in the solar neutrino problem's solution, however, have allowed determination of the sign of the small mass splitting, leaving two possible mass orderings that leave the second mass splitting undetermined: normal (NO) and inverted ordering (IO). Measuring the exact values of the three mass states and their ordering remains an open challenge in cosmology and particle physics. A related question is whether the mass values of the neutrinos are hierarchical (strongly distinct) or whether they are degenerate (nearly identical) in mass.  As discussed in the introduction, cosmological observations are very competitive to experimental bounds on neutrino mass as massive neutrinos influence both the expansion history and the growth of structure in the Universe. Recent years have witnessed increasingly stringent cosmological constraints on the sum of neutrino masses, ($\Sigma m_{\nu}$), approaching the lower bound permitted by the inverted hierarchy,  ($\sim 100\, \text{meV}$) \cite{DESI:2024mwx,Wang:2024hen}. This implies that the parameter space available for IO is becoming restricted. In this section, we will present the results of the analysis we developed to address these questions.
\subsection{Hierarchical Neutrino Mass Ordering}
Neutrino oscillation experiments have determined that the neutrino mass-squared differences for the solar and atmospheric oscillations are separated by about two orders of magnitude. Specifically, $\Delta m_{\rm solar}^2 =(7.53\pm0.18)\times 10^{-5}\,\mathrm{eV^2}$ and $\Delta m_{\rm atm}^2=(2.455\pm 0.028)\times 10^{-3}\,\mathrm{eV^2}$ [$\Delta m_{\rm atm}^2=(-2.529\pm 0.029)\times 10^{-3}\,\mathrm{eV^2}$] for the NO [IO] atmospheric case \cite{ParticleDataGroup:2024cfk}.
Consequently, there is also a minimum limit on the combined masses of the three active neutrinos ($\Sigma m_{\nu} = m_1+m_2+m_3$):
\begin{eqnarray*}
\Sigma m_{\nu}^{\rm{NO}} &=&
    m_1 
    + \sqrt{m_1^2 + \Delta m_{\rm solar}^2} 
    + \sqrt{m_1^2 + |\Delta m_{\rm atm}^2|} \, , 
    \\ \nonumber
\Sigma m_{\nu}^{\rm{IO}} &=&
    m_3 
    + \sqrt{m_3^2 + |\Delta m_{\rm atm}^2|} 
    \nonumber \\ &&
    + \sqrt{m_3^2 + |\Delta m_{\rm atm}^2| + \Delta m_{\rm solar}^2} \, ,
\end{eqnarray*}
where the lightest neutrino mass eigenstate is $m_1$ for NO and $m_3$ for IO. A key question from cosmology is whether NO or IO is preferred by the data in the case of a hierarchical neutrino mass spectrum. That is, hierarchy is where $m_1 \ll m_{2,3}$ (NO), or $m_3 \ll m_{1,2}$ (IO). Assuming the minimal mass case for both orderings, i.e. $m_1=0$ for NO and $m_3=0$ for IO, we obtain the minimal $\Sigma m_{\nu}$ for NO and IO by using the observed mass-splittings. This defines our two massive neutrino cases as: \textbf{NO} with $\Sigma m_{\nu} = 58\,\mathrm{meV}$ and \textbf{IO} with $\Sigma m_{\nu} = 101\,\mathrm{meV}$. 

At the hierarchical scale of interest for cosmological neutrino mass constraints, \textit{i.e.}, $m_1=0$ or $m_3=0$, it is inappropriate to model $\Sigma m_\nu$ as three degenerate mass states ($m_i=\Sigma m_\nu/3$, for $i=1,2,3$). Therefore, we model the massive neutrinos as one massive eigenstate ($m = \Sigma m_\nu$) and two massless eigenstates. Though it has been shown that this introduces some inaccuracies in the effects of massive neutrinos \cite{Lesgourgues:2004ps,Herold:2024nvk}, we show below that these effects do not affect our quantitative results appreciably, nor our conclusions. 

In the context of the $\Lambda$CDM model combined with the established determination of minimal neutrino masses from neutrino oscillation experiments, neutrinos have a minimum mass sum ($\Sigma m_{\nu}=58\,\mathrm{meV}$) that contribute to the matter density as in Eq.~\ref{eq:omnu}, which is determined by the standard thermal history. If cosmological results find $\Sigma m_{\nu}<58\,\mathrm{meV}$ with high significance, this would indicate physics beyond standard $\Lambda$CDM, its thermal history, or new physics in the neutrino sector. 
Given this minimum mass, cosmology tests NO versus IO as a binary choice in determining the neutrino's total mass scale. 

\begin{figure*}[h]
\centering
\includegraphics[width=\textwidth]{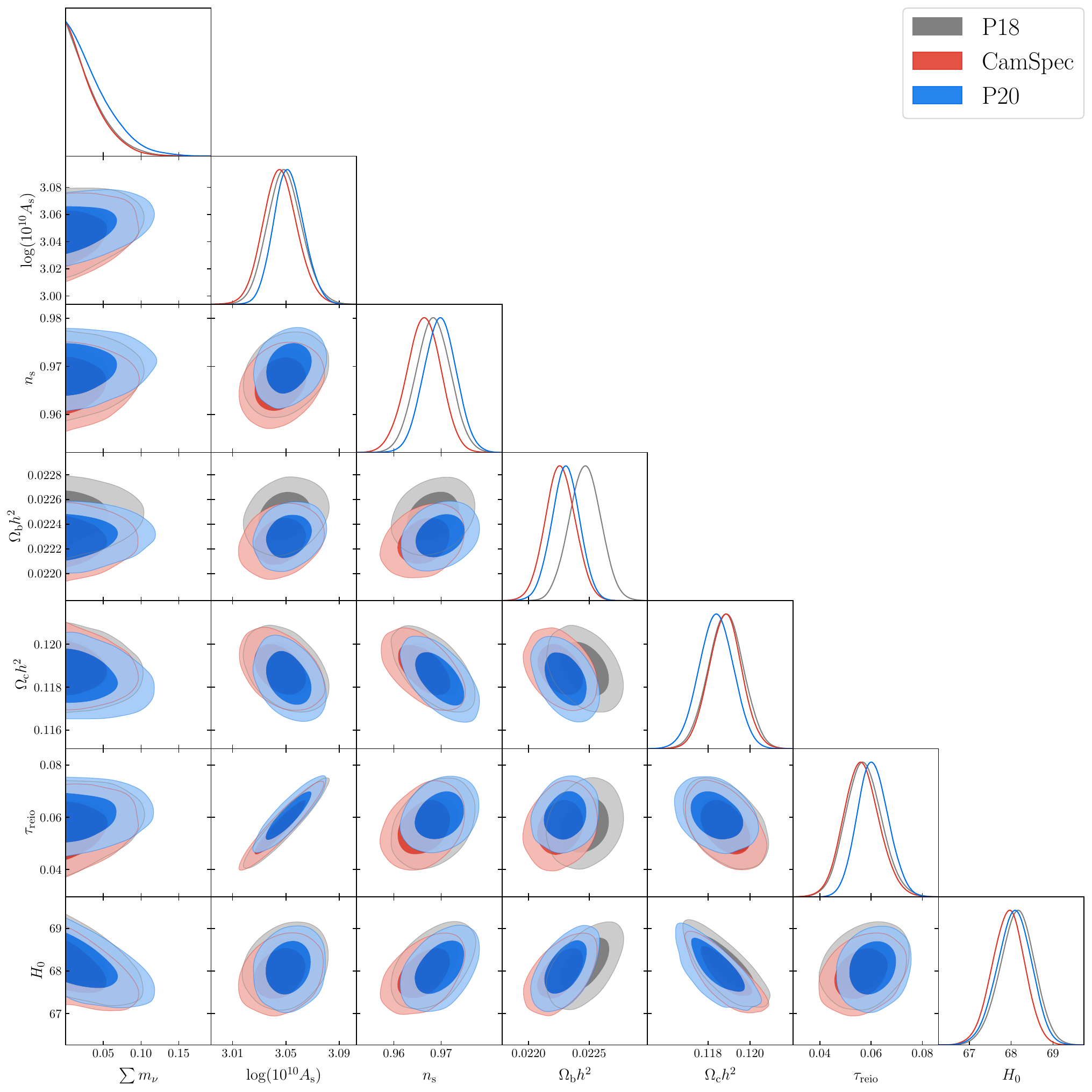}
\caption{Shown is the comparison of our benchmark datasets, but with different choices of the CMB analyses (P18, CamSpec, P20), with their 68\% and 95\% CL likelihoods of $\Lambda$CDM parameters plus $\Sigma m_{\nu}$.  $\Sigma m_{\nu}$ is in units of eV and $H_0$ in units of $\mathrm{km \,s^{−1}\,Mpc^{−1}}$.}
\label{fig:mnu_CMBs_PP}
\end{figure*}
To infer the cosmological data's preference for NO versus IO, we use the $\Delta \chi^2$ test, which is simply determined as the models have the same number of parameters with fixed neutrino masses (Eq.~\ref{eq:chi2}).  We calculate the preference of these scenarios including the benchmark likelihoods we defined in Section ~\ref{sec:lik}. We find preference for NO over IO of
\begin{equation}
    \Delta\chi^2(\text{NO over IO}) = 2.16\,,  %\left(1.47\sigma\right)\,,
\end{equation}
indicating a very weak preference, approximately $1.47\sigma$, for NO over IO.
Considering the preference of the value inconsistent with neutrino physics of $\Sigma m_\nu =0$, and we find \textit{non-standard} massless neutrinos preferred over NO by 
\begin{equation}
  \Delta\chi^2(\Sigma m_\nu =0\, \text{over NO})  = 1.85\,,  %\left(1.36 \sigma\right)\,,
\end{equation}
 indicating very weak preference, approximately $1.36\sigma$, of $\Sigma m_\nu =0$ over NO.
Importantly, the information on the matter content of the Universe from the supernovae surveys, which we adopt in our baseline cosmological dataset, slightly alleviates the cosmological constraint on neutrino mass. Without the PP SNe, we obtain $\Delta\chi^2(\text{NO over IO}) = 3.46\,  (1.86\sigma)$ and $\Delta\chi^2(\Sigma m_\nu =0\, \text{over NO}) = 2.36 \, (1.54\sigma)$. We find that the preference tests for the neutrino mass ordering are not substantially changed when adopting the CamSpec or Planck 2020 CMB data. To test the dependence on our default choice for modeling $\Sigma m_\nu$ as a single mass eigenstate, or as three degenerate states, we perform these $\chi^2$-tests with three degenerate neutrino masses, and find that the $\Delta\chi^2$ changes by less than 0.002 for both NO over IO and $\Sigma m_\nu =0$ over NO. Therefore, our results on evidence for mass ordering are not affected by our choice of modeling the hierarchy. Importantly, we find the sensitivity to neutrino mass ordering from P18+DESI+PP is about as strong as P18+Lensing+PP, showing that geometry is about as sensitive to neutrino mass as structure growth \cite{Loverde:2024nfi}.

\subsection{Free Neutrino Mass}
\label{sec:numass}
We use the several cosmological datasets available to infer the value $\Sigma m_{\nu}$ when it is allowed to vary freely. Our priors in the MCMC are shown in Table \ref{tab-priors}. We choose different combinations of cosmological likelihoods that we have presented in the introduction \S\ref{sec:lik} to test the effects of each cosmological epoch in our analysis. We use three sets of CMB data (P18, CamSpec and P20), CMB lensing, BAO DESI DR1, as well as the three SNe samples (PP, U3 and DES Y5).  In Fig. \ref{fig:1}, we present the 1D likelihood profiles of $\Sigma m_{\nu}$ for these cases via MCMC analysis. The respective choices of blue, yellow, and purple colored lines correspond to the posteriors resulting from adding PP, U3, and DES Y5 SNe samples.

 The nine different dataset combinations provide limits on the sum of the neutrino masses of 
\begin{align}
&\Sigma m_\nu (\text{P18, DESI, \& PP}) < 82.1\, \mathrm{meV} \, ,\nonumber\\ 
&\Sigma m_\nu (\text{P18, DESI, \& U3}) < 82.1\, \mathrm{meV} \, ,\nonumber\\
&\Sigma m_\nu (\text{P18, DESI, \& DES Y5}) < 98.0\, \mathrm{meV} \, ,\nonumber\\ 
&\Sigma m_\nu (\text{CamSpec, DESI, \& PP}) < 76.9\, \mathrm{meV} \, ,\nonumber\\ 
&\Sigma m_\nu (\text{CamSpec, DESI, \& U3}) < 77.0\, \mathrm{meV} \, ,\nonumber\\ 
&\Sigma m_\nu (\text{CamSpec, DESI, \& DES Y5}) < 86.6\, \mathrm{meV} \, ,\nonumber\\ 
&\Sigma m_\nu (\text{P20, DESI, \& PP }) < 94.1 \, \mathrm{meV} \, ,\nonumber\\ 
&\Sigma m_\nu (\text{P20, DESI, \& U3 }) < 93.8\, \mathrm{meV} \, ,\nonumber\\ 
&\Sigma m_\nu (\text{P20, DESI, \& DES Y5 }) < 108\, \mathrm{meV} \, ,
\end{align}
all at $95\%\,\mathrm{CL}$. We also test the effect of relaxing assumptions of the underlying cosmology from minimal $\Lambda$CDM's 6 parameters (plus neutrino mass) to include variation of curvature $\Omega_k$, evolving dark energy EoS, parameterized using  $w_0$ and $w_a$, and extra radiation $\Neff$, which is, in total, an 11 parameter model. In this case, the neutrino mass limit is only mildly relaxed, with 
\begin{equation}
    \Sigma m_\nu (\text{benchmark, 11 param.}) < 97.0\, \mathrm{meV}\,.
    \label{eq:11paramMnu}
\end{equation}
In Fig.~\ref{fig:1}, the black line corresponds to the 1D posterior distribution of the sum of the neutrino masses of an extended 11-parameter model with our benchmark dataset, P18, DESI, \& PP. We show the full inferred extended parameter values and errors in Table \ref{tab-priors}. 

\begin{figure*}[h]
\centering
\includegraphics[width=\textwidth]{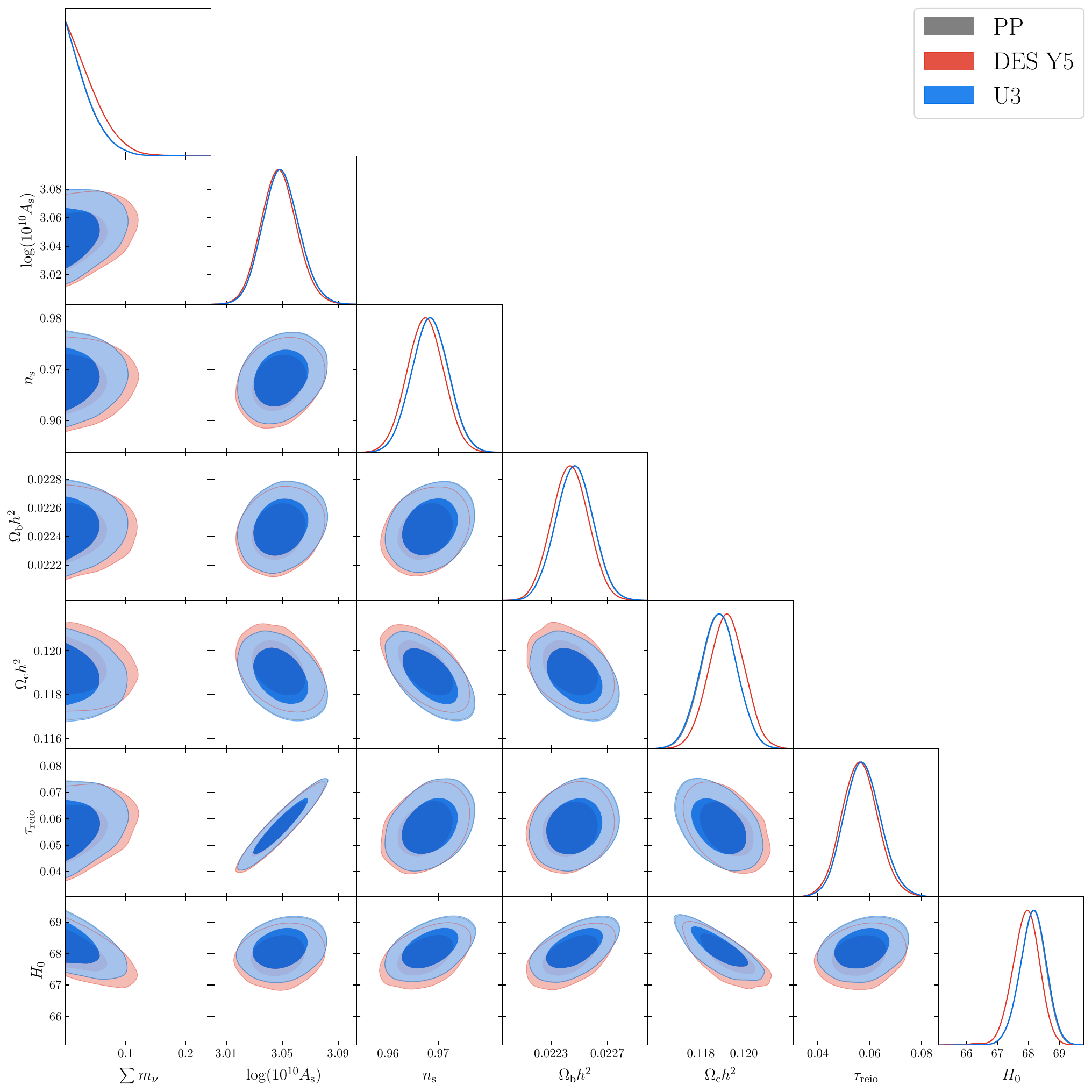}
\caption{Shown is the comparison of our benchmark datasets, but with different choices of the SNe samples (PP, DES Y5, and U3), with their 68\% and 95\% CL likelihoods of $\Lambda$CDM parameters plus $\Sigma m_{\nu}$.  Here, $\Sigma m_{\nu}$ is in units of eV and $H_0$ in units of $\mathrm{km \,s^{−1}\,Mpc^{−1}}$.}
\label{fig:mnu_SNes}
\end{figure*}

We observe three ``clusters'' of likelihoods in the $\Sigma m_{\nu}$ posteriors that we obtain and resulting limits, as shown in Fig.~\ref{fig:1}. The most stringent cluster of datasets includes five likelihoods: P18, CMB lensing, DESI, \& U3; P18, CMB lensing, DESI, \& PP; CamSpec, CMB lensing, DESI, \& U3; CamSpec, CMB lensing, DESI, \& PP; and, lastly, the 11 parameter chain with  P18, CMB lensing, DESI, \& PP. The moderate likelihood limit cluster includes four datasets: P18, CMB lensing, DESI, \& DES Y5; CamSpec, CMB lensing DESI, \& DES Y5; with P20, CMB lensing, DESI, \& U3; and P20, CMB lensing, DESI, \& PP. The weakest likelihood contour, in purple, is for P20, CMB lensing, DESI, \& DES Y5.

The constraints on mass are primarily influenced by two factors. First, variations in cosmological parameters arise from different primary CMB analyses. Among the three Planck CMB analyses, P20 demonstrates the highest consistency between the primary CMB data and Planck PR4 lensing, showing the smallest lensing anomaly. Consequently, the P20 CMB lensing signal is not significantly elevated, resulting in the weakest constraint on neutrino masses, consistent with what has been identified by other authors \cite{Allali:2024aiv,Naredo-Tuero:2024sgf}. 
CamSpec finds the most stringent constraints on $\Sigma m_{\nu}$, which occurs due to this analysis preferring a lower scalar spectral index, $n_s$, and lower optical depth to the last scattering surface, $\tau$, which both lower the amplitude of clustering on small scales, and are therefore positively correlated with $\Sigma m_{\nu}$ (see Fig.~\ref{fig:mnu_CMBs_PP}). As a result, their lower values drive a more stringent bound on $\Sigma m_{\nu}$.

Second, the neutrino mass limits are also sensitive to the adoption of which SNe dataset considered. As discussed in the introduction, neutrino mass is affected by CMB, BAO and SNe constraints in multiple ways. Of these 3 SNe samples, DES Y5 has the highest $\Omega_m$ which has two main impacts in cosmological observables: First, a higher $\Omega_m$ enhances growth, which can be offset by a higher $\Sigma m_\nu$. Secondly, the BAO scale is kept fixed when increasing $\Omega_m$ and decreasing $\Sigma m_\nu$, which goes in the opposite direction \cite{Thepsuriya:2014zda}. DES Y5 SNe sample differs greatly in its inference of cosmological parameters relative to the other two SNe samples, primarily with a much larger inferred $\Omega_m$ than the other SNe datasets \footnote{This may originate from a $\sim\! 0.04$ mag calibration offset between high and low redshifts \cite{Efstathiou:2024xcq,Dhawan:2024gqy}}. Combining all these effects, the main impact that drives the limit on $\Sigma m_{\nu}$ is the overall raise in $\Omega_m$ resulting on a relaxation on neutrino mass bounds when DES Y5 SNe sample is included. See Fig.~\ref{fig:mnu_SNes} for the parameter likelihood correlations for the three SNe datasets. Interestingly, the differences in the Planck CMB analyses are more significant in changes to cosmological parameters, including $\Sigma m_{\nu}$, than the SNe datasets.

There has been considerable interest recently on the tighter-than expected constraints on $\Sigma m_\nu$ to potentially indicate new physics, which could even be parameterized as ``negative'' neutrino mass by parameterizing the suppression of the growth of structure with massive neutrinos as enhanced growth when extending the growth of structure to be enhanced by the same linear suppression to linear enhancement of growth \cite{Craig:2024tky,Green:2024xbb,Naredo-Tuero:2024sgf,Elbers:2024sha,Allali:2024aiv}. To test the strength of the preference for ``negative'' neutrino mass, we fit Gaussian functions to the one-dimensional likelihoods of the chains for each dataset. We fit to all values of $[\Delta 2\ln({\mathcal{L}})]=2.5$ with respect to the likelihood maximum, as the MCMC process does not accurately sample the tails of the likelihood distributions. We recover the mean and standard deviation characteristic parameters of the associated Gaussian functions for the different datasets and calculate the deviation in $\sigma$ from the central value of the Gaussian to the neutrino physics minimal-mass NO case, i.e. $\Sigma m_{\nu}= 58\,\mathrm{meV}$. For all of the datasets we have analyzed where $\Sigma m_{\nu}$ is free, we find the deviation of the NO from the mean of the inferred likelihood to be between $1.13 \sigma$ and $1.97 \sigma$, always remaining below $2 \sigma$. 

\begin{figure}[h]
\centering
  \includegraphics[width=\columnwidth, height=0.73\columnwidth]{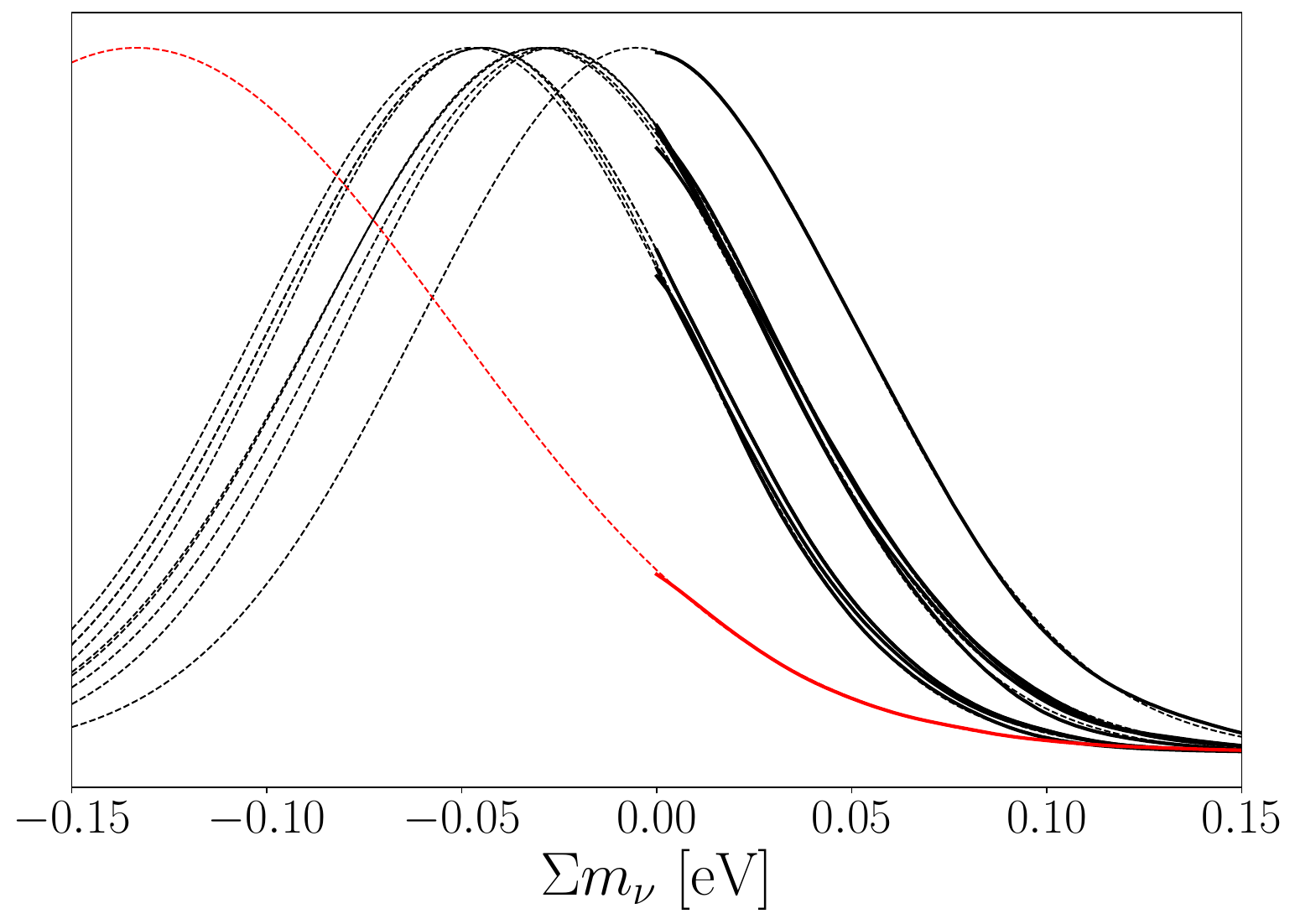}
   \caption{Shown here are the same 1D marginalized posterior distributions for $\Sigma m_{\nu}$ shown in Fig.~\ref{fig:1} (dashed lines), together with their Gaussian fits (solid lines). The black colored lines represent the different cosmological dataset combinations for the 7-parameter model where $\Sigma m_{\nu}$ can vary freely. The red lines correspond to the 11-parameter chain. Going from left to right in the peak of the black-colored likelihoods, the curves correspond to the cosmological likelihood combinations P18+DESI+U3; P18+DESI+PP; CamSpec\allowbreak+\allowbreak DESI\allowbreak+\allowbreak U3; CamSpec\allowbreak +\allowbreak DESI\allowbreak+\allowbreak PP; 
   P20+DESI+U3; P18+DESI+DES Y5, P20+DESI+PP; CamSpec+CMB lensing+DESI+DES Y5; and lastly the P20+DESI+DES Y5. The centroid for each combination, respectively, is below $2\sigma$ from the normal-ordering minimal neutrino mass, for all of the cases. The centroid for the red 11 parameter chain likelihood of the benchmark data is at the $2.31\sigma$ level. All datasets include CMB lensing.\label{fig:2} }
\end{figure}

The 11-parameter scenario, which offers the most model freedom of all the cases we study, finds a preference for the most negative value of inferred $\Sigma m_\nu$, with a tension level with respect NO of about $2.4\sigma$. We infer that this result is due to the much larger parameter space explored in the 11-parameter model, where neutrino masses are degenerate not only with $\Omega_m$, $A_s$, and $\tau$, but also with $w_0$, $w_a$, and $\Omega_k$. We present the 1D marginalized posterior distribution for $\Sigma m_\nu$ in this 11-parameter model in Fig.\ref{fig:2}.

\section{Exploring Beyond $\Lambda$CDM Models} 
\label{sec:model}
Despite $\Lambda$CDM's remarkable success in describing nearly everything about our Universe, there are some notable discrepancies that arise between observations and this theoretical framework that could indicate the presence of new physics beyond the standard model. Two of the most striking and long-lasting ones are the Hubble and $S_8$ tension. The Hubble tension is a significant, $\gtrsim\!5\sigma$, mismatch between the value of the expansion rate of the Universe today, which arises from a higher value based on the local expansion rate measured via a classic distance ladder, including Cepheid variables and Type Ia SNe, compared to a lower model-dependent value inferred using CMB observations and assuming $\Lambda$CDM. Another notable discrepancy involves the amplitude of matter fluctuations, $\sigma_8$, or the related parameter $S_8 = \sigma_8\sqrt{\Omega_m/0.3}$, which is inferred from large-scale structure surveys. The value derived assuming $\Lambda$CDM and using Planck CMB observations~\cite{Planck:2018vyg} is higher than those obtained through low-redshift probes such as weak lensing and galaxy clustering~\cite{Joudaki:2019pmv, Sugiyama:2023fzm, DES:2021bvc, DES:2021vln, DES:2021wwk,Miyatake:2023njf,KiDS:2021opn}.

In this section, we present our study of candidate extensions to the $\Lambda$CDM paradigm.  We perform minimization of the models we introduce in this section using the benchmark likelihoods we presented in Sec.~\ref{sec:lik}. We evaluate preference for extensions to $\Lambda$CDM with the Bayesian evidence provided by the AIC, discussed in \S\ref{sec:intro}, and level of tension in $\sigma$ with respect to $\Lambda$CDM.  We test extensions by comparing the cosmological datasets described above, on their own, and with the addition of external $H_0$ and $S_8$ constraint likelihoods, both separately and together.

\subsection{$\Lambda$CDM+$N_{\textrm{eff}}$}
\label{sec:Neff}
\begin{figure}[t!]
\centering
\includegraphics[width=\columnwidth]{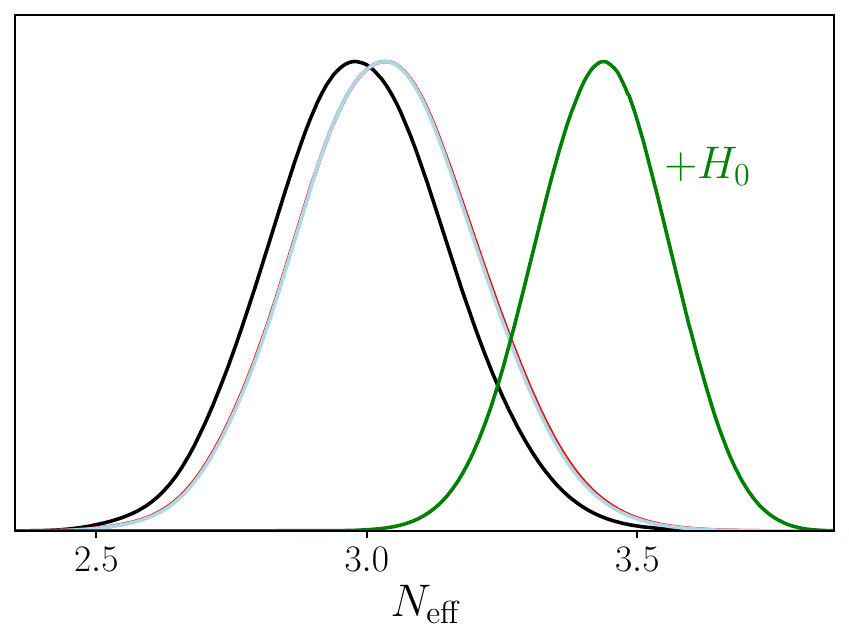}\caption{\label{fig:nnu} Shown here is a comparison of 1D marginalized posterior distributions for $N_{\rm eff}$  for different SNe data combinations when added to the P18, CMB lensing and DESI likelihoods. The solid dark blue, dashed light-blue, and black lines correspond to the PP, U3, DES Y5 SNe samples respectively, with all consistent with $N_\nu^\textrm{SM}$. The green line is the resulting posterior distribution when the SH0ES $H_0$ measurement likelihood is added to the PP chain. finding $N_{\rm eff}= 3.45 \pm 0.12$.}  
\end{figure}

Standard $\Lambda$CDM and the particle content of the Standard Model of particle physics predicts the relativistic energy density of the early Universe to be consistent with $N_{\textrm{eff}} = N_\nu^\textrm{SM} = 3.044$. The calculations of $N_\nu^\textrm{SM}$ take into account neutrino oscillations, incomplete neutrino decoupling through $e^\pm$ annihilation, and QED corrections \cite{Akita:2020szl,Bennett:2020zkv,Froustey:2020mcq}. The possibility of extra radiation density is prevalent in Beyond the Standard Model (BSM) extensions of particle physics \cite{DiValentino:2020zio}. This has gained extra attention as additional relativistic energy density has also been shown to alleviate the Hubble tension \cite{DiValentino:2021izs}. The possibility of lower relativistic energy density than $N_\nu^\textrm{SM}$ is also possible in low-reheating temperature universes or other nonstandard thermal histories. A high-$\Neff$ model exemplifies scenarios where the energy density of additional components, like Early Dark Energy or the presence of BSM particles, plays a significant role in the drag epoch early Universe. Adding extra relativistic particles beyond standard model neutrinos, such as dark radiation,  reduces the sound horizon, resulting in the sound horizon's greater compatibility with a higher $H_0$. This additional energy density influences the positions of CMB acoustic peaks relative to the photon damping scale, which are precisely measured by CMB observations. Therefore, the alteration of the sound horizon by an increased $\Neff$ to allow for compatibility with the local $H_0$ limit comes into tension with the independent constraints on $\Neff$ from the CMB acoustic peaks and damping scale. 

We concretely specify the cosmology of a varying $\Neff$ as one where the active neutrinos are exactly one species with relativistic energy density of $N_\nu = 1$, and $\Sigma m_\nu = 58\,\mathrm{meV}$. The remainder of $\Neff$ is pure relativistic energy density (modeled in \texttt{CAMB} as \texttt{num\_nu\_massless}). An interesting nonstandard scenario that can be parameterized with an additional contribution of massless ultra-relativistic species is a low reheating temperature cosmology \cite{Kawasaki:1999na}. Our results can be interpreted in that framework (for all values of $\Neff$), or as an effective early extra relativistic energy density from any BSM relativistic component (for cases where $\Neff > N_\nu^\textrm{SM}$). 

We first determine a preferred $N_{\textrm{eff}}$ with our baseline datasets. We run a 7-parameter extended model allowing $N_{\textrm{eff}}$ to vary freely between the priors reported on Table \ref{tab-priors}, using MCMC chains including P18+CMB lensing+DESI BAO DR1 and the three different SNe samples. We find that the inferred relativistic energy density is identical for both the PP and Union3 datasets, 
\begin{equation}\Neff = 3.05\pm 0.17\,,\end{equation}
shown as the red line and light blue lines in Fig. \ref{fig:nnu}. For DES Y5 SNe, we find 
\begin{equation}\Neff  = 2.99 \pm 0.17\,,\end{equation} 
shown as the black line in Fig. \ref{fig:nnu}. 

When adding the SH0ES $H_0$ measurement, we find 
\begin{equation}N_{\rm eff}=3.45 \pm 0.12\,. \end{equation}
Significantly, we find that this non-standard $\Neff$ model is strongly favored over $\Lambda$CDM on the Jeffreys' scale, with a $\Delta \text{AIC}=-10.9$. The 1D marginalized posterior distributions for $N_{\rm eff}$ when the SHOES $H_0$ measurement's likelihood is added to our benchmark likelihood set is shown by the green curve in Fig.~\ref{fig:nnu}. This increased energy density in $\Neff$ would be in tension with with (BBN) constraints \cite{Cyburt:2015mya}. However, several mechanisms exist for making this $\Neff$ compatible with BBN \cite{Abazajian:2002bj}. For example, if a considerable lepton asymmetry is present in the Universe, $\Neff \approx 3.5$ is compatible with the light element abundances \cite{Burns:2022hkq}.

When only the DES Y3 $S_8$ measurement is included, a free $\Neff$ model  does not provide a better fit with respect to $\Lambda$CDM, with a
$\Delta\text{AIC}=2.3$ therefore indicating $\Neff$ is not the driving physics for a solution to both $H_0$ and $S_8$ tensions, a result found previously. When we include the measurements of DES Y3 $S_8$ and SH0ES $H_0$, $\Neff$+$\Lambda$CDM remains strongly favored over $\Lambda$CDM, with 
$\Delta\text{AIC}=-8.0$. However, this result is driven by the $H_0$ measurement and is not indicative of a preference for non-standard $\Neff$ from $S_8$.

\subsection{Sterile Neutrinos}

Anomalies in short-baseline neutrino experiments may be due to the existence of one or more sterile neutrinos, e.g., the LSND, MiniBooNE, reactor, and gallium anomalies \cite{Conrad:2013mka,Dasgupta:2021ies,Norman:2024hki}. 
Sterile neutrinos are significantly constrained by cosmology~\cite{Hamann:2011ge,Planck:2018vyg} and other neutrino experiments~\cite{Maltoni:2002xd,DayaBay:2016lkk}. However, there remains much interest on sterile neutrinos' impact on cosmology and their potential signals ~\cite{Hasegawa:2020ctq,Gariazzo:2013gua,Archidiacono:2014apa,Abazajian:2017tcc}. Reconciling eV sterile neutrinos with cosmology often involves new mechanisms that suppress their full thermalization  \cite{Hasegawa:2020ctq,Jacques:2013xr}. 

The possibility of $\Neff \approx 3.5$ indicated by the Hubble tension could be due to partially thermalized sterile neutrinos. In addition, massive sterile neutrinos could be responsible for the suppression of $S_8$ \cite{Wyman:2013lza}, though massive sterile neutrinos are constrained by the same physics as the active neutrinos in \S\ref{sec:numass}. Therefore, we examine here how eV-scale sterile neutrinos affect cosmological tensions like $H_0$ and $S_8$.

In this model, we consider the following relativistic species: photons,  3 active and one additional sterile neutrino.  
In our work, we have fixed the active neutrino sector to give a contribution of $N_\nu^{\textrm{SM}} = 3.044$ to $N_{\textrm{eff}}$, with two massless and one massive neutrino with a mass of 58 meV. Therefore, the contribution to $N_{\textrm{eff}}$ from the sterile species is simply $\Delta N_{\textrm{eff}} = N_\nu^{\textrm{SM}} - 3.044$. When the sterile neutrino is relativistic at early times, assuming the only radiation species are photons and neutrinos, the contribution of a light sterile neutrino to $N_{\textrm{eff}}$ is given by \cite{Acero:2008rh},
\begin{equation}
\Delta N_{\textrm{eff}} =\left[\frac{7}{8}\frac{\pi^2}{15}T_{\nu}^{4} \right]^{-1} \frac{1}{\pi^2} \int dp ~p^3 ~f_s (p),
\end{equation}
where $T_\nu$ is active neutrino temperature, $p$ is the neutrino momentum, and $f_s(p)$ is momentum distribution function of the sterile neutrino. At late times its energy density is parametrized as an effective mass \cite{Acero:2008rh,Planck:2013oqw}:
\begin{equation}
\omega_s \equiv \Omega_s h^2 = \frac{m_s^{\textrm{eff}}}{94.1 \textrm{eV}} = \frac{h^2 m_s^{\textrm{ph}}}{\pi^2 \rho_{c}} \int dp~ p^2 ~f_s (p),
\end{equation}
where $\rho_{c}$ is the critical density, $\Omega_s h^2$ is the sterile neutrino energy density. Since sterile neutrinos do not have electroweak interactions and mix, through oscillations, with the active neutrinos, they are not necessarily ever in thermal equilibrium. Active neutrinos decouple at a temperature $T \sim 1$ MeV, they are relativistic. Hence $f_s (p)$ does not depend on the physical mass of the sterile neutrino, $m_\mathrm{s}^{\textrm{ph}}$. However, $f_s (p)$ depends on the production mechanism of the light sterile neutrino. 

We consider non-thermal production of sterile neutrinos. One such mechanism is the Dodelson-Widrow (DW) process \cite{Dodelson:1993je}, for which $f_s (p) = \beta (e^{p/T_{\nu}}+1)^{-1}$, where $\beta$ is a normalization factor. 
We investigate the possibility that the massive sterile neutrinos are either partially or fully thermalized. 

\subsubsection{Partially Thermalized}
%Partially thermalized sterile neutrinos 
A partially thermalized sterile neutrino model requires two additional free parameters  to the six $\Lambda$CDM ones:
\begin{equation}
m^{\textrm{eff}}_{\textrm{s}} = \Delta N_{\textrm{eff}}~ m_{\textrm{s}}^\mathrm{ph};~~~~~~~~\Delta N_{\textrm{eff}} =\beta.
\end{equation}
Here, $m^{\textrm{eff}}_{\textrm{s}}$ is \textit{not the physical mass}. When this parameter and $\beta$ are allowed to vary freely one encounters the regime where $\beta \sim 0.01$ and the physical mass $m_{\textrm{s}}^\mathrm{ph}\sim 1\,\mathrm{keV}$, \textit{i.e.} the warm dark matter regime. To focus on the eV scale, we specified cases where $m^{\textrm{eff}}_{\textrm{s}}$ and $\beta$ are plausible in short-baseline-motivated eV-scale models. We focus on three thermalization scenarios for each of either a 1 eV or 0.1 eV sterile neutrino, from partial thermalization cases from Ref.~\cite{Hasegawa:2020ctq} (their Fig.~5).  The three scenarios have values of $\Delta N_{\rm eff} = 0.45 , 0.074,\, \&\, 0.0086$, corresponding to $N_{\rm eff} = 3.49, 3.12,\, \&\, 3.05$. These values are for low reheating temperature models where $T_\textrm{RH}=4 \mathrm{MeV}$ for mixing angles of $\sin^2 2\theta = 0.001, 0.01, 0.1$ respectively, but can be achieved in other partial-thermalization cases \cite{Jacques:2013xr}. As discussed in \S\ref{sec:Neff}, a high $\Neff$ can be consistent with BBN and the light element abundances with a number of physical mechanisms, including a nontrivial chemical potential in the electron neutrinos \cite{Orito:2000zb,Abazajian:2002bj,Burns:2022hkq}, which can also suppress or enhance sterile neutrino thermalization \cite{Abazajian:2004aj,Mirizzi:2012we}.

We study the three cases of partial thermalization of neutrinos, using our benchmark datasets of P18, CMB lensing, DESI DR1, PP, plus SH0ES $H_0$. Interestingly, our results find the existence of a thermalization threshold preference. For a 1 eV partially thermalized sterile neutrino with 
$N_{\rm eff} = 3.49, 3.12,\, \&\, 3.05$, 
adding the $H_0$ measurement likelihood, we find $\Delta \mathrm{AIC} = 25.9, 0.8, 1.1$, respectively, showing that below a certain thermalization limit, between the first and second cases, this model transitions from being strongly disfavored by cosmological data to being at a similar tension level when compared to $\Lambda$CDM. Note the $\Delta \mathrm{AIC}$ values for all the partially thermalized neutrino models we analyze are identical with the models' $\Delta \chi^2$ preferences since no additional free parameters are introduced.  

Even more significantly, our results find that a 0.1 eV partially-thermalized sterile neutrino is \textit{more favored} than $\Lambda$CDM, when adopting the SH0ES $H_0$ measurement.  In the three thermalization scenarios we considered, \textit{i.e.} with  $N_{\rm eff} = 3.49, 3.12,\, \& \, 3.05$, when including the SH0ES $H_0$ measurement, we find $\Delta \mathrm{AIC} = -11.0, -3.7, \,\& \, -2.1$. All of these cases are preferentially fit by our baseline data, alleviating the Hubble tension more effectively than $\Lambda$CDM. Note that the model with and 0.1 eV partially-thermalized sterile neutrino with $N_{\rm eff} = 3.49$ is strongly preferred over $\Lambda$CDM by its AIC, as well as its $\Delta \chi^2$, at $>\!3\sigma$. Its preference is comparable to and very slightly better than the pure relativistic energy $\Neff$ studied in \S\ref{sec:Neff}. This preference is in part due to the degenerate correlation between $\Neff$ and $\Sigma m_\nu$ in the BAO scale, which allows for both to increase without changing the BAO scale \cite{BOSS:2014hhw}.
\begin{figure}[h!]
\centering
\includegraphics[width=\columnwidth]{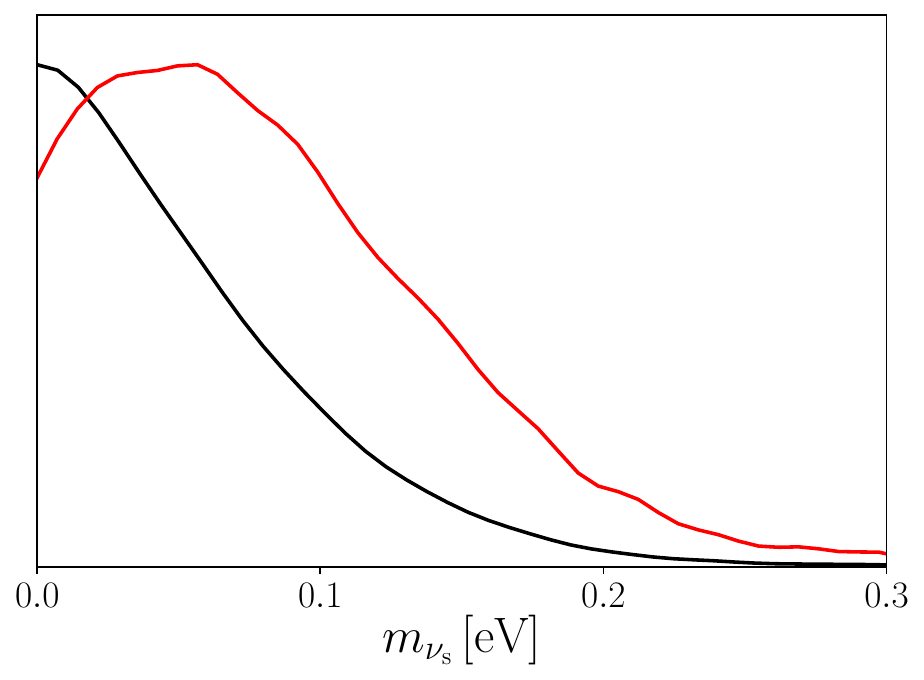}
   \caption{\label{fig:fullthe} Comparison of the one-dimensional marginalized posterior distributions for $m_{\nu_{\rm s}}$ is shown, for the case of a fully thermalized sterile neutrino ($\Neff=4.044$). The black curve represents the benchmark data alone, while the red curve includes the SH0ES $H_0$ measurement. }
\end{figure}

\subsubsection{Fully Thermalized}

We explore the possibility of a fully thermalized sterile neutrino in addition to the three active neutrinos setting $\beta = 1$ and $\Delta N_{\rm eff} = 1$, so $N_{\rm eff}= 4.044$. This model introduces only one additional parameter $m^{\textrm{eff}}_{\textrm{s}}$ which is the physical mass. In this model, we fix $N_{\rm eff}= 4.044$ and vary $m^{\textrm{eff}}_{\textrm{s}}$ together with the other 6 $\Lambda$CDM parameters. With our benchmark data, of P18, CMB lensing, DESI DR1, \& PP, a fully thermalized sterile neutrino, with free particle mass, is strongly disfavored compared to $\Lambda$CDM with $\Delta \mathrm{AIC} > 20$. In addition, despite a thermalized sterile neutrino possibly lowering clustering amplitudes and accommodating the DES Y3 $S_8$, this is strongly disfavored by the baseline data, at the level of $\Delta \mathrm{AIC} > 20$.

Interestingly, when including our baseline datasets with the SH0ES $H_0$ measurement, we find $\Delta\mathrm{AIC} = 4.5$, not strongly disfavored with respect to $\Lambda$CDM. Remarkably, this means that a fully-thermalized eV-scale short-baseline sterile neutrino is roughly as consistent with the $H_0$ tension as $\Lambda$CDM is.  Fig.~\ref{fig:fullthe} shows the 1D marginalized posterior distribution resulting from running the MCMC associated with this fully thermalized model for the \textit{physical mass} $m_{\nu_{\rm s}}$ for the benchmark data alone (black), and with the SH0ES $H_0$ measurement (red). 
%%%%%%%%%%%%%%%%%%%%%%%%%%%%%%%%%%%%%%%%%%%%%%%%%%%%%%%%%%%%%%%%%%%%%%%%%%%%%%%%%%%%%%%%%%%%%%%%

\begin{figure*}[]
\centering
\includegraphics[width=\textwidth]{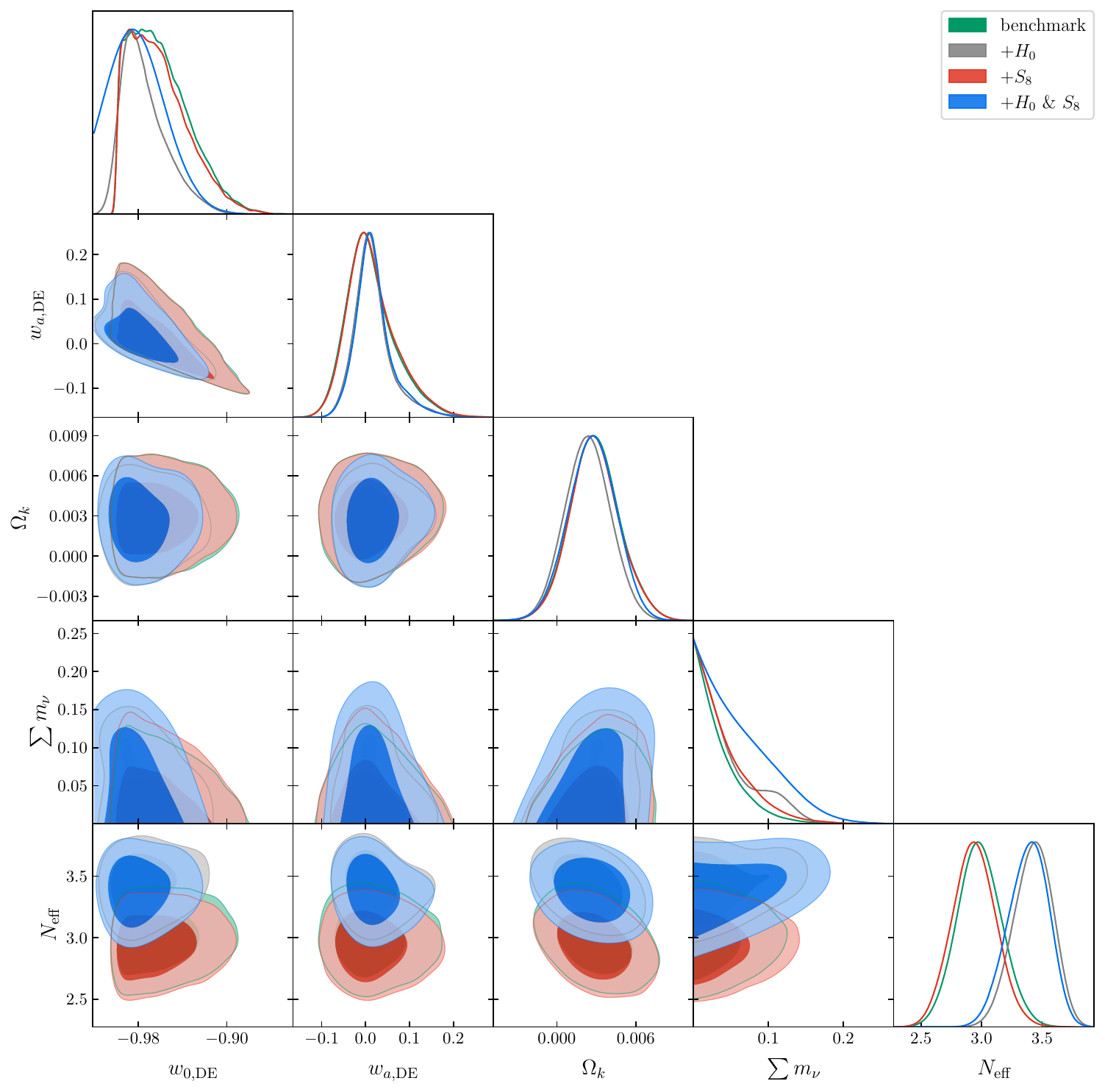}
\caption{Shown are the 68\% and 95\% CL likelihood contours of the extension parameters in our 11-parameter case, using our benchmark data. Without the tension data, no significant shifts are seen in any extended parameters, except for $w_0= −0.975^{+0.012}_{-0.024}$. As with more economical models, only $\Neff$ deviates strongly from its standard value when including $H_0$. Here, $\Sigma m_{\nu}$ is in units of eV and $H_0$ is in units of $\mathrm{km \,s^{−1}\,Mpc^{−1}}$.}
\label{fig:mnu_11param}
\end{figure*}

%%%%%%%%%%%%%%%%%%%%%%%%%%%%%%%%%%%%%%%%%%%%%%%%%%%%
%%%%%%%%%%%%%%%%%%%%%%%%%%%%%%%%%%%%%%%%%%%%%%%%%%%%

%%%%%%%%%%%%%%%%%%%%%%%%%%%%%%%%%%%%%%%%%%%%%%%%%%%%%%%%%%%%%%%%%%%%%%%%%%%%%%%%%%%%%%%%%%%%%%%%
\begin{table*}[]
\centering
\setlength{\tabcolsep}{3pt} % Adjust column spacing
\begin{tabularx}{\textwidth}{|X|>{\centering\arraybackslash}X|>{\centering\arraybackslash}X|>{\centering\arraybackslash}X|>{\centering\arraybackslash}X|>{\centering\arraybackslash}X|}
\hline
%\begin{tabularx}{\textwidth}{|X|X|X|X|X|X|}
\textbf{Parameters} & \textbf{Benchmark} & \textbf{Benchmark} & \textbf{+ $H_0$ Posterior} & \textbf{+ $S_8$ Posterior} & \textbf{+ $H_0$ \& $S_8$ Posteriors } \\ \hline
$\Omega_\mathrm{b} h^2$        & $0.02247 \pm 0.00013$ & $0.02236 \pm 0.00022$ &  $0.02274^{+0.00017}_{-0.00020}$                      &              $0.02237 \pm 0.00022$       &          $0.02273^{+0.00018}_{-0.00021}$             \\ \hline
$\Omega_\mathrm{c} h^2$        & $0.1185 \pm 0.0083$  & $0.1184 \pm 0.0028$   &  $0.1251^{+0.0028}_{-0.0025}$        &           $0.11756 \pm 0.0027$           &      $0.1241^{+0.0029}_{-0.0025}$                \\ \hline
$\log(10^{10} A_\mathrm{s})$    & $3.049 \pm 0.013$  & $3.053 \pm 0.016$     &  $3.073 \pm 0.018$                    &       $3.049 \pm 0.016$               &           $3.072 \pm 0.015$          \\ \hline
$n_\mathrm{s}$               & $0.9683 \pm 0.0036$    & $0.9639 \pm 0.0084$   &   $0.9808 ^{+0.0076}_{-0.0069}$                   &    $0.9636 \pm 0.0084$                  &       $0.9804 ^{+0.0076}_{-0.0068}$               \\ \hline
$\tau_{\rm reio}$             & $0.0586 \pm 0.0071 $  & $0.0586^{+0.0069}_{-0.0077}$ &            $0.0616 \pm 0.0089$  &        $0.0581 \pm 0.0074$              &     $0.0619^{+0.0074}_{-0.0060}$                 \\ \hline
$100\theta_\mathrm{MC}$       & $1.04108 \pm 0.00029$  & $1.04107 \pm 0.00043$ &      $1.04044 \pm 0.00036$                &           $1.04119 \pm 0.00043$           &       $1.04048 ^{+0.0038}_{-0.0043}$               \\ \hline
$\Sigma m_{\nu}\ (95\%\ CL)$            &  $< 82.1$\,meV    & $< 97.0$\,meV  &      $< 125$\,meV                 &       $< 112$\,meV        & $< 130$\,meV         \\ \hline
$N_{\rm eff}$             &       & $2.98 \pm 0.18$       &                   $3.43 \pm 0.15$   &        $2.94 \pm 0.18$              &  $3.39^{+0.18}_{-0.13}$                    \\ \hline
$\Omega_k$                &       & $0.0029 \pm 0.0019$   &                  $0.0023 \pm 0.0017$    &         $0.0029  ^{+0.0017}_{-0.0020}$           &   $0.0027 \pm 0.0017$                    \\ \hline
$w_0$                    &        & $-0.961^{+0.012}_{-0.037}$ &  $-0.975^{+0.012}_{-0.024}$                &     $-0.962^{+0.012}_{-0.036}$                 &         $-0.978^{+0.018}_{-0.031}$             \\ \hline
$w_a$                   &         & $0.013^{+0.039}_{-0.066}$ &                $0.016^{+0.027}_{-0.045}$  &      $0.015^{+0.039}_{-0.060}$                &          $0.019^{+0.028}_{-0.043}$            \\ \hline
\end{tabularx}
\caption{Best-fit values and $68\%$ confidence intervals for the parameters are presented for the 7- and 11-parameter extended models are shown. Our benchmark data includes P18, CMB lensing, DESI, and PP likelihoods, both without external data and with the addition of $H_0$ and $S_8$ posteriors, considered individually and in combination.}
\label{tab:posteriors}
\end{table*}
\subsection{The $S_8$ Tension and Massive Active or Sterile Neutrinos}

A reduction in the clustering amplitude on smaller scales, leading to lower values of $\sigma_8$ or $S_8$ within $\Lambda$CDM, can be achieved by increasing the neutrino mass and its contribution to the overall density of matter~\cite{Hu:1997mj}. Several papers have interpreted low $S_8$ measurements as possible evidence for nonzero active neutrino masses or combinations involving additional mass eigenstates \cite{Battye:2013xqa,Wyman:2013lza,BOSS:2014etx}. However, the strong constraints from the baseline datasets disfavor massive neutrinos at or above the 0.1 eV scale, as discussed in \S\ref{sec:numass}. In addition, massive neutrinos do not alter the physics of the local expansion rate relative to the CMB-epoch sound horizon. In our $\Delta \chi^2$ and AIC tests, including both the $H_0$ and $S_8$ tension measurements disfavors massive active neutrinos, as well as either partially or fully thermalized massive sterile neutrinos, relative to $\Lambda$CDM, though at a level lower than the strong threshold.

\subsection{Evolving Dark Energy, Massive Neutrinos, Extra Relativistic Energy Density, and Curvature}
Single-parameter modifications studied in the Planck 2018 analysis~\cite{Planck:2018vyg} often fail to reconcile the $H_0$ and $S_8$ tensions. For example, introducing changes in parameters such as the sum of neutrino masses $\Sigma m_\nu$, spatial curvature $\Omega_k$, or the effective number of neutrino species $N_\mathrm{eff}$ affects $H_0$ and $S_8$ in similar directions, limiting their ability to resolve both tensions simultaneously \cite{Escudero:2022rbq}.
To test whether multi-parameter simultaneous modifications of dark energy, neutrino physics, relativistic energy density and/or geometry may be responsible for the observed $H_0$ and $S_8$ tensions, we vary all of these model components in our 11-parameter model. 
We relax the standard $\Lambda$CDM assumptions and use a model which allows the six parameters of the standard $\Lambda$CDM model to vary freely, along with five additional parameters: $N_{\rm eff}, \Sigma m_{\nu}, \Omega_k, w_0$ and $w_a$.

\textit{Dynamical Dark Energy (DDE)} --- 
A time-dependent EoS for dark energy would be a major discovery, and is indicated at almost $4\sigma$ by the DESI BAO data, when combined with SNe data \cite{DESI:2024mwx}. This evolution may also affect other cosmological tensions. While simple variations in dark energy EoS alone might not fully resolve the Hubble tension, they could improve model fits when combined with sterile neutrinos or other parameters~\cite{Keeley:2019esp}. During the era where dark energy prevails, phantom dark energy with an EoS $w < -1$ induces an accelerated expansion of the Universe more pronounced than in $\Lambda$CDM with $w = -1$. This allows for the $H_0$ determined from CMB experiments to align better with the $H_0$ derived from local Universe measurements~\cite{DiValentino:2016hlg,Ludwick:2017tox,Vagnozzi:2019ezj,Alestas:2020mvb}. There are several parameterizations for $w$ that can be examined to study this model \cite{DiValentino:2017zyq,Yang:2018qmz}. In this work we will use the Chevallier-Polarski-Linder (CPL) parameterization~\cite{Chevallier:2000qy, Linder:2002et}, where the dark energy EoS is given by $w(a) = w_0 + w_a(1 - a)$. This parameterization is advantageous due to its simplicity \cite{Linder:2002et}; however, interpreting constraints in the $w_0$ \& $w_a$ space in terms of transitions to phantom-like behavior (where the EoS $w < -1$) requires caution, since observational data only constrain the EoS over a limited range of redshifts, restricting the ability to fully capture high and low redshift dependencies. 
The behavior of $w(z)$ outside this constrained redshift range stems from the parameterization itself rather than from the data, indicating that this parameterization should be regarded as purely phenomenological. Recent findings from the DESI collaboration have used this parameterization and have been interpreted to suggest the existence of evolving dark energy \cite{DESI:2024mwx}. Particularly, the data show a preference for a present-day quintessence-like EoS ($w_0 > -1$) that crossed the phantom barrier ($w_a < 0$).  However, this interpretation is  sensitive to the Type Ia SNe sample used in conjunction with DESI's measurements of BAO, when combined with CMB observations \cite{Efstathiou:2024xcq}. Their results find a preference that reaches a significance level of about $3.5\sigma$ and $3.9\sigma$ when combining P18, DESI BAO with Union3 and DES Y5, respectively, while it is reduced to around $2.5\sigma$ with PP~\cite{DESI:2024mwx,Giare:2024ocw}.

\textit{Nonzero Curvature} ---
The angular diameter distance, measured through low-redshift observations such as BAO and SNe, is directly related to the Universe's curvature. Permitting a non-flat Universe by treating the curvature density parameter, $\Omega_k$, as a variable introduces an additional degree of freedom, which influences the geometry of spacetime. 
Nonzero curvature also has implications for structure formation, offering a potential approach to alleviate the $S_8$ tension. Such models with nonzero curvature can be consistent with low-redshift observations, as explored in Refs.~\cite{Ryan:2018aif,Handley:2019tkm,DiValentino:2019qzk,Cao:2021ldv}. 

The best-fit values of this 11 parameter model, when we include the benchmark data, P18, CMB Lensing, BAO and PP likelihoods, are shown in Table~\ref{tab:posteriors}. The parameters for this extended model, in the first column of Table~\ref{tab:posteriors}, are remarkably consistent with $\Lambda$CDM, with the exception of $w=-0.961^{+0.012}_{-0.037}$, which is more positive than $\Lambda$ by over 3$\sigma$. The curvature $\Omega_k = 0.0029 \pm 0.0019$ is less than 2$\sigma$ away from flat. Notably, evidence for a nontrivial $w_a$ is not present, which is likely due to the less significant changes away from $\Lambda$CDM in the other nonstandard parameters, $\Neff$ and $\Omega_k$, shown in the green contours in Fig. \ref{fig:mnu_11param}.

\textit{Tension Data} --- We test to see if there is indication of new physics in the extended model when including the $H_0$ and $S_8$ tension data. We show the changes in the extension parameters when the $H_0$ and $S_8$ measurement likelihoods are included in Fig.~\ref{fig:mnu_11param}. The primary change observed with the addition of the Hubble constant measurement is a slightly more negative central value for $w_0= −0.975^{+0.012}_{-0.024}$, with a smaller lower bound error. Additionally, as with the 7 parameter $\Lambda$CDM plus $\Neff$ model, there is a significant increase in the relativistic energy density, $N_{\rm eff} = 3.43 \pm 0.15$. There is a relaxation of the neutrino mass bound, with $\Sigma m_{\nu} <125 \, \textrm{meV}\ (95\% \, \textrm{C.L.})$. Importantly, when adding curvature, neutrino mass, and $\Neff$, there is no evidence for evolution of the dark energy equation of state, $w_a=0.016^{+0.027}_{-0.045}$. Curvature does not significantly shift, with $\Omega_k = 0.0023 \pm 0.0017$. 
    
When the $S_8$ measurement is included, the main difference is a relaxation of the neutrino mass upper bound to $\Sigma m_{\nu} <112 \,  \textrm{meV}\ (95\% \, \textrm{C.L.})$. The relativistic energy density is consistent with the standard value, $N_{\rm eff} = 2.94 \pm 0.18$. The remaining parameters $w_0$, $w_a$ and $\Omega_k$ do not suffer notable deviations.

\section{Discussion \& Conclusions}
\label{sec:concl}

Our analysis studies several current cosmological survey datasets with respect to a minimal $\Lambda$CDM model, with specific attention to the inferred sum of neutrino masses, effective neutrino number, and sterile neutrino models. 
One of the fundamental questions hoped to be answered by cosmology is whether a normal or inverted ordering is preferred in the hierarchical neutrino mass case, where the lightest neutrino mass is negligible relative to the other two mass eigenstates. We include SNe survey data to complement constraints on cosmological parameters, using a suite of nine sets of CMB (P18, CamSpec, P20) and SNe (PP, Union3, DES Y5) data, along with the DESI DR1 BAO data. We propose and use the robust $\Delta \chi^2$-test in determining the cosmological evidence for NO or IO.  In this hierarchical case, using our benchmark data, we find that the preference for NO over IO remains weak, $1.47\sigma$, with similar preferences for all 9 dataset contributions that we explored. When we exclude SNe data, the preference for NO over IO becomes very slightly stronger, at $1.86\sigma$. Similarly, the cosmological data slightly favor massless neutrinos, at $1.36\sigma$, in our benchmark data, with similar preferences in all 9 datasets. Though matter clustering modeling can be affected \cite{Lesgourgues:2004ps,Herold:2024nvk} by the choice of a single or three degenerate neutrino mass eigenstates---which are both incorrect at the minimal mass hierarchy, when two neutrinos are massive---we find our conclusions do not change when altering the selection of the number of nonzero neutrino masses.

Exploring neutrino masses as free, we find limits on their sum in a range between $\Sigma m_\nu < 76.9\, \mathrm{meV}$ up to $\Sigma m_\nu < 108\, \mathrm{meV}$ $(95\%\,\mathrm{CL})$. As pointed out in other work, the strong constraints on $\Sigma m_\nu$ arise in large part to the lensing anomaly, which indicates stronger amplitude CMB lensing power than expected \cite{Craig:2024tky,Green:2024xbb}. 
The variation in the limits on $\Sigma m_\nu$ arise from the separate groups' analyses of the Planck CMB data and the separate SNe datasets. As has been known for some time, $\Omega_m$ is positively correlated with $\Sigma m_\nu$ in the amplitude of clustering of matter, as measured by CMB lensing. Conversely, $\Omega_m$ and $\Sigma m_\nu$ are largely negatively correlated in the BAO scale. As discussed in Ref.~\cite{Loverde:2024nfi}, the measurements of the BAO scale distance ratios, as well as distances from SNe, are as sensitive of a probe of neutrino mass as the growth of structure, and sometimes more so.

The range of mass limits are affected by two primary factors: First, the cosmological parameters vary as inferred by the separate primary CMB analyses. Between the three CMB analyses, P18, CamSpec, and P20, P20 has the most consistency between the primary CMB and Planck PR4 lensing, and therefore the least lensing anomaly. Therefore, the P20 CMB lensing signal is not significantly anomalously high, and therefore the P20 inferred limit on neutrino masses is the weakest. Second, the matter density inferred by the SNe samples differs, as parameterized by $\Omega_m$ or $\Omega_c h^2$, with DES Y5 data preferring higher $\Omega_m$ than either PP or U3. This higher value of $\Omega_m$ required by DES Y5 shifts the likelihood in the direction of the positive correlation of $\Omega_m$ with the CMB lensing amplitude. The commensurate positive correlation of $\Omega_m$ with $\Sigma m_\nu$ results in a higher inferred $\Sigma m_\nu$, for the DES Y5 case. These effects are illustrated in the triangle plots in Fig. \ref{fig:mnu_CMBs_PP} and Fig. \ref{fig:mnu_SNes}, which show the correlations between $\Sigma m_\nu$, $A_s$, $n_s$, $\Omega_b h^2$, $\Omega_c h^2$, $\tau$ and $H_0$ cosmological parameters across the different datasets.

As more cosmological survey data have become available, the constraints on the sum of neutrino masses generally have become progressively tighter, with current 95\% CL upper limits crossing below the minimum value permitted by the inverted mass hierarchy, $\Sigma m_\nu = 101\,\mathrm{meV}$. However, as we discuss above, the preference for NO over IO remains weak. We analyze the shape of the likelihoods for $\Sigma m_\nu$, and we find a shape that has a peak to the inferred likelihood be at a negative sum of neutrino masses, as reported in previous work \cite{Craig:2024tky,Green:2024xbb,Naredo-Tuero:2024sgf,Elbers:2024sha,Jiang:2024viw}. In our results, the extrapolated negative central value of the inferred $\Sigma m_{\nu}$ relative to the minimum required by NO remains below $2 \sigma$ significance. The preference for ``negative'' neutrino mass is primarily from the same physics that is responsible for very tight $\Sigma m_{\nu}$ limits discussed above, namely, the high amplitude of the CMB lensing signal disfavoring any suppression of the matter power spectrum due to neutrino mass.

We explore the $H_0$ and $S_8$ tension datasets. As found in previous work, extra relativistic energy density, $N_{\rm eff}=3.45 \pm 0.12$ significantly alleviates the $H_0$ tension. Note, a high $\Neff$ can be accommodated by BBN and the light element abundances by a number of physical mechanisms, including a nontrivial chemical potential in the electron neutrinos  \cite{Orito:2000zb,Abazajian:2002bj,Abazajian:2004aj,Burns:2022hkq,Mirizzi:2012we}.

Anomalies in short-baseline neutrino experiments have indicated the possible existence of light sterile neutrinos \cite{Dasgupta:2021ies}. Very significantly, our results indicate a preference for a 0.1 eV partially-thermalized sterile neutrino at $3.3\sigma$ relative to standard $\Lambda$CDM when including the $H_0$ measurement. This preference is particularly strong for scenarios with approximately $N_{\rm eff}\approx 3.5$, suggesting that partially thermalized sterile neutrinos consistent with the short-baseline signals could provide a viable explanation for the Hubble tension. Fully thermalized sterile neutrinos are comparable to $\Lambda$CDM in fitting the benchmark cosmological data when the $H_0$ measurement is included. Excluding the $H_0$ data, fully or partially thermalized sterile neutrinos remain disfavored. No extended model we test alleviates the $S_8$ tension.

We also studied an extended 11-parameter model that includes nontrivial curvature, $\Neff$, and evolving dark energy via $w_0\,\&\,w_a$. The cosmology inferred by the extended model does not significantly differ in its parameters from those in the $\Lambda$CDM framework, except for $w_0= −0.975^{+0.012}_{-0.024}$, which has about $2\sigma$ preference for being non-$\Lambda$. In this extended model, the neutrino mass limit is not very significantly weakened,  with $\Sigma m_\nu < 97.0\, \mathrm{meV}$ (95\% CL). 

Cosmology is forecast to achieve high-sensitivity to neutrino mass and number, with upcoming data from ongoing CMB analyses from the ACT \cite{ACT:2020frw} and SPT \cite{SPT:2021efh} experiments, as well as cosmological structure surveys DESI and Euclid. The CMB-S4 experiment, combined with current and ongoing experiment datasets, should achieve $5\sigma$ sensitivity to NO, and a sensitivity to extra relativistic degrees of freedom of $\sigma(\Neff) \approx 0.03$ \cite{CMB-S4:2016ple,Brinckmann:2020bcn}. These new data will set the stage for unprecedented precision in cosmology that could lead to the long-anticipated cosmological detection of neutrino mass or, if undetected, signal new physics beyond the standard model for cosmology or particle physics \cite{CMB-S4:2016ple, Euclid:2024yrr}.

\begin{acknowledgments}
HGE and KNA would like to thank discussions with Ryan Keeley and Jui-Lin Kuo during the initial stages of this work, and Antony Lewis for helpful discussions on implementation of sterile neutrinos in \texttt{CAMB}. HGE thanks Shouvik Roy Choudhury for useful discussions about the sterile neutrino implementation in Cobaya. We thank Marc Kamionkowski, Marilena Loverde, and Zach Weiner for detailed comments on the manuscript. KNA is partially supported by the U.S. National Science Foundation (NSF) Theoretical Physics Program Grant No.\ PHY-2210283. 
\end{acknowledgments}

%-------------------------------------------------------------------------
\bibliography{references}

%-------------------------------------------------------------------------------
\end{document}